\documentclass[11pt, conference, onecolumn, compsocconf]{IEEEtran}

\usepackage{dt,amssymb,bbm,amsmath,listings}
\usepackage[ruled,vlined,linesnumbered,shortend]{algorithm2e}
\SetKwInOut{Input}{input}
\SetKwInOut{Requires}{requires\hspace*{.5mm}}
\SetKwInOut{Returns}{returns}
\SetKwInOut{InOutBlank}{$\;$}
\SetKw{Procedure}{procedure}

\newlength\breite

\newenvironment{ylist}[1]{%
  \settowidth\breite{#1}\begin{list}{#1}{%
      \setlength\itemsep{0pt}\setlength\topsep{.7ex}\setlength\parsep{.7ex}%
      \setlength\itemindent{0mm}\setlength\listparindent{0mm}\setlength\labelsep{1ex}%
      \setlength\labelwidth{\breite}%
      \addtolength\breite{1ex}\addtolength\breite{.7ex}\setlength\leftmargin{\breite}%
      }}{\end{list}}

\begin{document}
\newtheorem{theorem}{Theorem}
\newtheorem{example}{Example}
\newtheorem{Pro}[theorem]{Proposition}
\newtheorem{corollary}[theorem]{Corollary}
\newtheorem{lemma}[theorem]{Lemma}
\newcommand{\angl}[1]{\langle #1 \rangle}
\newcommand{\sem}[1]{[\![ #1 ]\!]}
\newcommand{\rsem}[1]{[ #1 ]}
\newcommand{\DT}{\textsc{dt}}
\newcommand{\XML}{\textsc{xml}}
\newcommand{\MSO}{\textsc{mso}}
\newcommand{\BUTA}{\textsc{dbta}}
\newcommand{\DTR}{\textsc{dt}^{\text{R}}}
\newcommand{\DMTT}{\textsc{dmt}}
\newcommand{\DTTA}{\textsc{dtta}}
\newcommand{\dtta}{\textsc{dtta}}
\newcommand{\HDToL}{\textsc{hdt0l}}
\newcommand{\DMTTnsn}{\textsc{dmt}_{\text{nsn}}}
\newcommand{\DMTTRnsn}{\textsc{dmt}_{\text{nsn}}^{\text{R}}}
\newcommand{\DMTTR}{\textsc{dmt}^{\text{R}}}
\newcommand{\rhs}{\mathsf{rhs}}
\newcommand{\dom}{\mathsf{dom}}
\newcommand{\makeAff}{\mathsf{makeAff}}
\newcommand{\notInAff}{\mathsf{notInAff}}
\newcommand{\mr}{\mathsf{mr}}
\newcommand{\bin}{\mathsf{bin}}
\newcommand{\ts}{t}
\newcommand{\ws}{w}
\newcommand{\bt}{\mathbf{t}}
\newcommand{\bw}{\mathbf{w}}
\newcommand{\bp}{\mathbf{p}}
\newcommand{\bz}{\mathbf{z}}
\newcommand{\bx}{\mathbf{x}}

\def\proofend{\hfill\IEEEQEDclosed}
\newcommand{\qed}{\proofend}
\def\RHS{\text{{\scshape rhs}}}
\def\Nat{\mathbb N}
\def\T{{\cal T}}
\def\L{{\cal L}}
\def\Q{{\mathbb Q}}




\title{Equivalence of Deterministic Top-Down Tree-to-String Transducers is Decidable}


\author{\IEEEauthorblockN{Helmut Seidl}
	\IEEEauthorblockA{Fakult\"at f\"ur Informatik\\
			TU M\"unchen\\
	Garching, Germany\\
	Email: seidl@in.tum.de}
	\and
	\IEEEauthorblockN{Sebastian Maneth}
	\IEEEauthorblockA{School of Informatics\\
		University of Edinburgh\\
        Edinburgh, UK\\
	Email: smaneth@inf.ed.ac.uk}
	\and
	\IEEEauthorblockN{Gregor Kemper}
	\IEEEauthorblockA{Fakult\"at f\"ur Mathematik\\
			TU M\"unchen\\
	Garching, Germany\\
	Email: kemper@ma.tum.de}
	}

\maketitle

\begin{abstract}
We show that equivalence of deterministic top-down tree-to-string transducers
is decidable, thus solving a long standing open problem in formal language theory.
We also present efficient algorithms for subclasses:
polynomial time for total transducers with unary output alphabet 
(over a given top-down regular domain language), 
and co-randomized polynomial time for linear transducers;
these results are obtained using techniques from multi-linear algebra.
For our main result, we prove that equivalence can be certified by means of 
inductive invariants using polynomial ideals.
This allows us to construct two semi-algorithms, one searching for a
proof of equivalence, one for a witness of non-equivalence.
Furthermore, we extend our result to deterministic top-down tree-to-string transducers which produce
output not in a free monoid but in a free group.
\end{abstract}











\newcommand{\blank}{\textvisiblespace}
%


\section{Introduction}\label{s:intro}

\noindent
Transformations of structured data are at the heart of functional 
programming~\cite{Wadler90,MarlowW92,VoigtlanderK04,Voigtlander05,DBLP:conf/pepm/MatsudaIN12} 
and also application areas such as compiling~\cite{FV98},
document processing~\cite{XSLT,xquery,Maneth1999,DBLP:journals/acta/EngelfrietM03,DBLP:conf/pods/ManethBPS05,DBLP:conf/icdt/ManethPS07,DBLP:conf/icde/HakutaMNI14}, 
automatic translation of natural languages~\cite{liu2006tree,liu2007forest,maletti2009power,DBLP:conf/acl/BrauneSQM13} 
or even cryptographic protocols~\cite{KustersW07}.
The most fundamental model of such transformations is given by 
(finite-state tree) \emph{transducers}~\cite{man03,FV98}.
Transducers traverse the input by means of finitely many mutually recursive 
functions --- corresponding to finitely many \emph{states}. Accordingly, these transducers are also 
known as \emph{top-down} tree transducers~\cite{DBLP:journals/jcss/Thatcher70,DBLP:journals/mst/Rounds70} 
or, if
additional parameters are used for accumulating output values,
\emph{macro} tree transducers~\cite{EV85}.
Here we only deal with deterministic transducers and denote them $\DT$ and $\DMTT$ transducers, 
respectively
(equivalence is undecidable already for very restricted classes of non-deterministic transducers~\cite{DBLP:journals/jacm/Griffiths68}).

When the output is produced in a structured way, i.e., in the case of tree-to-tree transducers,
many properties, at least of transducers without parameters, are fairly well understood.
An algorithm for deciding equivalence of $\DT$ transducers already dates back to the 80s~\cite{Esik81}.
More recently, canonical forms have been provided allowing for effective minimization~\cite{EngelfrietMS09}
as well as Gold-style learning of transformations from examples~\cite{LemayMN10}.
%
%
In various applications, though, the output is \emph{not} generated in a structured way.
This may be the case when general scripting languages are employed~\cite{JensenMM11},
non-tree operations are required~\cite{persei04} or simply, because the result is a string.

Assume, e.g., that we want to generate a 
well-formed {\sc xml} document from an internal tree-like
representation where the elements of the document do not only have tags and contents but also 
attributes. The output for the input tree
\[
\begin{array}{l}
{\sf frame}(	\\
\qquad{\sf defs} ({\sf height}(20), {\sf defs}(	
		  {\sf width}(50),{\sf end})),	\\
\qquad{\sf content}({\sf button}(\text{"Do\blank not\blank press!"}),\ldots)	\\
)
\end{array}
\]
then should look like:
\[
\begin{array}{l}
\langle{\sf frame}\blank{\sf height}=\text{\sf "}20\text{\sf "}\blank{\sf width}=\text{\sf "}50\text{\sf "}\rangle\\
\qquad\langle{\sf button}\rangle \text{Do\blank not\blank press!}\langle{\sf /button}\rangle	\\
\qquad \ldots	\\
\langle{\sf /frame}\rangle
\end{array}
\]
This translation could be accomplished by a tree-to-string transducer
with, among others, the following rules:
\[
\begin{array}{rll}
q({\sf frame}(x_1,x_2))&\to&
		\langle{\sf frame}~q_1(x_1) q(x_2)\langle{\sf /frame}\rangle	\\
q_1({\sf end})	&\to& \rangle	\\
q_1({\sf defs}(x_1,x_2))&\to& q_2(x_1) q_1(x_2)	\\
q_2({\sf height}(x_1))	&\to& \blank{\sf height} = \text{\sf "}q_3(x_1)\text{\sf "}	\\
q_2({\sf width}(x_1))	&\to& \blank{\sf width} = \text{\sf "}q_3(x_1)\text{\sf "}	\\
q({\sf button}(x_1)) 	&\to&
		\langle{\sf button}\rangle q_3(x_1)\langle{\sf /button}\rangle 	\\
	&\ldots&	\\
\end{array}
\]
According to the peculiarities of {\sc xml}, arbitrary many attribute value pairs may be positioned inside the start tag
of an element. The given rules generate the closing bracket of the start tag from the node ${\sf end}$
which terminates the list of attribute definitions. 
At the expense of an empty right-hand side, the closing bracket could also be generated by the rule for
the tag ${\sf frame}$ directly. In this case, the two first rules should be replaced with:
\[
\begin{array}{rll}
q({\sf frame}(x_1,x_2))&\to&
                \langle{\sf frame}~q_1(x_1)\rangle q(x_2)\langle{\sf /frame}\rangle   \\
q_1({\sf end}) &\to& \epsilon       \\
\end{array}
\]
while all remaining rules stay the same.
This example indicates that already simple tasks for structured data may be solved by
transducers processing their inputs in different ways.

Following the tradition since~\cite{Eng80}, we denote 
\emph{tree-to-string transducers} by prefixing the letter $y$ 
(which stands for ``yield'', i.e., the function that turns a tree into the
string of its leaf labels, read from left to right).
In~\cite{Maneth1999,persei04} dedicated transducers for \XML{} have been introduced.
Beyond the usual operations on trees, these transducers also support concatenation of outputs.
Decidability of equivalence for these transducers has been open.
Since they can be simulated by tree-to-string transducers, our main result implies that
equivalence is decidable for both types of transducers (where for~\cite{persei04}
we mean the parameter-less version of their transducers).

Amazingly little has been known so far for tree-to-string transducers, 
possibly due to the multitude of ways in which they can produce the same output string.
As a second example, consider the transducer $M$ with initial state $q$, on input trees 
with a ternary symbol $f$ and a leaf symbol $a$, defined by:
\[
\begin{array}{lcl}
q(f(x_1,x_2,x_3))&\to &q(x_3)a q_1(x_2)b q(x_2)\\
q_1(f(x_1,x_2,x_3))&\to &q_1(x_3)q_1(x_2)q_1(x_2) ba\\
q_1(e)&\to & ba\\
q(e)&\to & ab.\\
\end{array}
\]
Furthermore, let $M'$ be the transducer with single state $q'$ and the rules:
\[
\begin{array}{lcl}
q'(f(x_1,x_2,x_3))&\to &ab q'(x_2)q'(x_2)q'(x_3)\\
q'(e)&\to & ab.\\
\end{array}
\]
Some thought reveals that the transducers $M$ and $M'$ are equivalent, although the output
is generated in a quite ``un-aligned'' way with respect to $x_2,x_3$. 
Note that these two transducers do not fall into any class of tree-to-string transducers
for which equivalence has been known to be decidable so far.
%
%
One class where equivalence is already known to be decidable, are the
\emph{linear and order-preserving} deterministic tree-to-string transducers 
as studied in~\cite{StaworkoLLN09}.
A transducer is linear, if each input variable $x_i$ occurs at
most once in every right-hand side. A transducer is order-preserving
if the variables $x_i$ appear in ascending order (from left to right)
in each right-hand side.
Equivalence for these can be decided by 
a reduction to Plandowski's result~\cite{DBLP:conf/esa/Plandowski94} even
in polynomial time~\cite{StaworkoLLN09}.
This class of transducers is sufficiently well-behaved so that
periodicity and commutation arguments over the output
strings can be applied to derive canonical normal forms~\cite{LaurenceLNST11} and enable
Gold-style learning~\cite{LaurenceLNST14}.
Apart from these stronger normal form results, equivalence itself
can indeed be solved for a much larger class of tree-to-string translations,
namely for those definable in \MSO{} logic~\cite{engman06}, or equivalently, by
macro tree-to-string translations of \emph{linear size increase}~\cite{engman03}.
This proof 
gracefully uses Parikh's theorem and the theory of semi-linear sets. 
More precisely, for a Parikh language $L$ (this means $L$,
if the order of symbols is ignored, is equivalent to a regular language)
it is decidable whether there exists an output string with equal number of $a$'s and $b$'s
(for given letters $a\not= b$). The idea of the proof is to construct $L$
which contains $a^nb^m$ if and only if, on input $t$,
transducer $M_1$ outputs $a$ at position $n$ and
transducer $M_2$ outputs $b$ at position $m$.

Our main result generalizes the result of \cite{engman06} by proving that equivalence of \emph{unrestricted} deterministic top-down 
tree-to-string transducers is decidable. 
By that, it solves an intriguing problem which has been open for at least
thirty-five years~\cite{Eng80}. The difficulty of the problem may perhaps become apparent as
it encompasses not only the equivalence problem for \MSO{} definable transductions, but also
the famous \HDToL\ sequence equivalence problem~\cite{culkar86,ruh86,DBLP:journals/tcs/Honkala00a}, 
the latter is the sub-case when the input is restricted to monadic trees~\cite{DBLP:journals/corr/Maneth14}.
Opposed to the attempts, e.g., in~\cite{StaworkoLLN09}, we refrain from any arguments based on the
combinatorial structure of finite state devices or output strings. We also do not follow the line of arguments 
in~\cite{engman06} based on semi-linear sets. 
Instead, we proceed in two stages. In the first stage, we consider transducers with unary output alphabets
only (Sections~\ref{s:total}--\ref{s:general}). In this case, a produced output string can be represented 
by its length, thus turning the transducers effectively into tree-to-integer transducers. 
For a given input tree, the output behavior of the states of such a unary $y\DT$ transducer
is collected into a vector. 
Interestingly, the output vector for an input tree turns out to be a \emph{multi-affine} 
transformation of the corresponding output vectors of the input subtrees. 
As the property we are interested in can be
expressed as an affine equality to be satisfied by output vectors, we succeed in replacing the \emph{sets} of reachable
output vectors of the transducer by their \emph{affine closures}. 
This observation allows us to apply exact fixpoint techniques 
as known from abstract interpretation of programs~\cite{Muller-OlmS04a},
to effectively compute these affine closures and thus to decide equivalence.
%
In the next step, we generalize these techniques to a larger class of transducers, namely, 
unary \emph{non-self-nested} transducers. These are transducers which additionally have parameters, but
may use these only in a restricted way. Although they are more expressive than unary $y\DT$ transducers,
 the effect
of the transducer for for each input symbol still is multi-affine and therefore allows a similar (yet more expensive)
construction as for unary $y\DT$ transducers for deciding equivalence.
In the final step, we ultimately show that the restriction of \emph{non-self-nestedness} can be lifted. 
Then however, multi-affinity is no longer available. 
In order to attack this problem, we turn it upside down:
instead of maintaining affine spaces generated by sets of input trees, we maintain their \emph{dual},
namely suitable properties satisfied by input trees. Indeed, we show that \emph{inductive invariants}
based on conjunctions of polynomial equalities
are sufficient for proving equivalence. 
This result is obtained by representing conjunctions of equalities by polynomial ideals~\cite{weiss}
and applying Hilbert's basis theorem to prove that finite conjunctions suffice. 
We also require compatibility of polynomial ideals with Cartesian products, a property which may be of 
independent interest.
For the specific case of monadic input alphabets, we obtain a decision procedure which resembles 
techniques for
effectively proving polynomial program invariants by means of weakest pre-conditions 
\cite{Letichevsky93,Muller-OlmS04}. 
For non-monadic input symbols, we obtain two semi-decision procedures,
one enumerating all potential proofs of equivalence, while the other searches for counter-examples.

Having established decidability for all $y\DMTT$ transducers with unary output alphabet, 
we indicate in the second stage how these transducers are able to simulate $y\DT$ transducers 
with multiple output symbols
when these are viewed as \emph{digits} in a suitable number system
(Section~\ref{s:mtt}).
The corresponding construction maps \emph{linear} $y\DT$ transducers 
into unary non-self-nested $y\DMTT$ transducers, 
and general $y\DT$ transducers into general unary $y\DMTT$ transducers.
In this way, our algorithms for deciding equivalence of unary transducers give rise to 
algorithms for deciding equivalence of linear and general $y\DT$ transducers, respectively.
While decidability of (in-)equivalence for linear $y\DT$ transducers has been known 
(via the result of \cite{engman06}),
the resulting randomized polynomial complexity bounds has only recently been improved to 
a truly polynomial upper bound by Boiret and Palenta \cite{DBLP:conf/dlt/BoiretP16}. 
No other method, on the other hand, allows to decide equivalence of unrestricted $y\DT$ transducers.

\subsection*{Related Work}\label{ss:related}

\noindent
Decision procedures for equivalence of deterministic tree transducers have been provided for various
sub-classes of transducers (see, e.g., \cite{DBLP:journals/corr/Maneth14} for a recent survey). 
The equivalence problem for $y\DT$ transducers 
has already been mentioned as a difficult open question in \cite{Eng80}.
Still, little progress on the question has been made for tree transducers where the outputs are
unstructured strings.
The strongest result known so far is the decidability of equivalence for
MSO definable tree-to-string transductions \cite{engman06}; this class 
is equal to $y\DMTT$ transducers of linear size increase \cite{engman03}.
For the specific sub-class of
linear $y\DT$ transducers, 
we obtain an algorithm with by far better complexity bounds as those provided by 
the construction in \cite{engman06}.
General
MSO definable tree-to-string transductions on the other hand,
can be simulated by $y\DT$ transducers with regular look-ahead
(see \cite{DBLP:journals/iandc/EngelfrietM99,DBLP:journals/jcss/EngelfrietM02}).
Since equivalence of $y\DT$ transducers with regular look-ahead can be reduced to equivalence 
of $y\DT$ transducers without
look-ahead but relative to a $\DTTA$ automaton, 
our decidability result encompasses the decidability result
for MSO definable tree-to-string transductions.
It is more general, since
$y\DT$ transducers may have more than linear size increase and thus may not be MSO definable. 
The same holds true for arbitrary $y\DMTT$ transducers with unary output alphabet.
%
%
It is unclear how or whether the suggested methods can be generalized to the equivalence problem of 
unrestricted $y\DT$ transducers.

The methods employed to obtain our novel results have the following predecessors.
The algorithm for deciding equivalence in the case of non-self-nested $y\DMTT$ transducers is related to
the algorithm in~\cite{Seidl90} for deciding \emph{ambiguity} equivalence of non-deterministic finite tree automata.
Two automata are ambiguity equivalent if they agree 
for each input tree, in the numbers of accepting runs.
While vector spaces and multi-linear mappings were sufficient in case of finite automata, we required
\emph{affine} spaces and \emph{multi-affine} mappings in case of linear $y\DT$ transducers.

The known algorithm for deciding equivalence of $y\DT$ transducers with \emph{monadic} alphabet is based on a reduction to
the \HDToL\ sequence equivalence problem. The latter can be solved~\cite{DBLP:journals/tcs/Honkala00a}
via establishing an increasing chain of finite sets of word equations 
which are guaranteed to eventually agree in their sets of solutions.
Instead, our elegant \emph{direct} algorithm for monadic unary $y\DMTT$ transducers is related to an algorithm 
effective program verification. In \cite{Letichevsky93,Muller-OlmS04} a decision procedure is presented
which allows to check whether a given polynomial equality is invariably true at
a given program point of a polynomial program, i.e., a program with non-deterministic branching
and polynomial assignments of numerical variables. Similar to the new algorithm for every state $p$ of the
$\DTTA$ automaton, the verification algorithm characterizes the required conjunction
of polynomial equalities at each program point by polynomial ideals. These ideals then are characterized
as the least solution of a set of constraints which closely resembles those in equation \eqref{eq:wp}.
To the best of our knowledge,
the algorithm for solving the equivalence problem in the unrestricted case, is completely new.

\subsection*{Organization of the remaining paper}

After providing basic notations and concepts in the next section, we first concentrate on 
unary transducers. In Section~\ref{s:total}, we provide an algorithm for deciding equivalence
of $y\DT$ transducers with unary output alphabet. In Section~\ref{s:nonnested}, we generalize
this algorithm to non-self-nested $y\DMTT$ transducers with unary output alphabet. These methods are
based on least fixpoint iterations over affine spaces.
The case of $y\DMTT$ transducers with unary output alphabet and arbitrary nesting is considered in
Section~\ref{s:general}. This section makes use of polynomial ideals in a non-trivial way.
As kind of warm-up, a dedicated algorithm for $y\DMTT$ transducers 
with monadic input alphabet is provided
which is based on \emph{least} fixpoint iteration over polynomial ideals.
The general method for arbitrary input alphabets goes beyond that.
It is based on the notion of \emph{inductive invariants} which provide proofs of equivalence.
The strongest inductive invariant, however, can only be characterized non-effectively via 
a \emph{greatest fixpoint} over polynomial ideals. 
Section~\ref{s:general} provides us with a decision procedure of equivalence based on 
two semi-algorithms. In-equivalence can easily be verified by providing a witness inputs 
for which the outputs differ. Searching for a proof of equivalence, on the other hand,
may be rather difficult.
Therefore, Section~\ref{s:algo} provides a systematic means to identify candidate inductive
invariants.
Section~\ref{s:mtt} indicates how the algorithms for unary $y\DT$ transducers can be used to
decide equivalence for $y\DT$ transducers with arbitrary output alphabets by providing an appropriate
simulation.
Section~\ref{s:free} then shows that equivalence of $y\DT$ transducers remains
decidable if the output is not considered as a string, i.e., an element of the \emph{free monoid},
but as an element in the \emph{free group}. This means that we now allow symbols to have positive or
negative polarities and assume that symbols with opposite polarities may cancel each other
out. In order to deal with this situation, we consider a setting where the output symbols
of a $y\DT$ transducer are interpreted as square matrices with rational entries and show that
in this setting, equivalence is still decidable. By recalling that the free group with
two generators is a sub-group of the special linear group of $(2\times 2)$ matrices
with entries in $\mathbb Z$, we thus arrive at the desired result.
Finally, Section~\ref{s:results} discusses applications of the obtained results to
various models of transducers as proposed in the literature.

A preliminary version of this paper was presented at FOCS'2015~\cite{DBLP:conf/focs/SeidlMK15}.
That version has been extended with a practical algorithm for enumerating inductive invariants 
(Section~\ref{s:algo}) and by allowing output not just in the free monoid of strings, but
instead in the free group (Section \ref{s:free}).

\section{Preliminaries}\label{s:prelim}

\noindent
For a finite set $S$, we denote by $|S|$ its cardinality.
For $m\in\Nat$ let $[m]$ denote the set $\{1,\dots,m\}$.
A ranked alphabet $\Sigma$ is a finite set of symbols each 
with an associated natural number called rank.
By $a^{(m)}$ we denote that $a$ is of rank $m$ and 
by $\Sigma^{(m)}$ the set of symbols in $\Sigma$ of rank~$m$. 
For an $m$-tuple $\bt$ and $i\in[m]$ we denote 
by $\bt_i$ the $i$th component of $\bt$.
The set $\T_\Sigma$ of trees over $\Sigma$ 
is the smallest set $T$ such that if
$\bt\in T^m$ for $m\geq 0$ then also
$f\,\bt\in T$ for all $f\in\Sigma^{(m)}$. 
For the tree $f()$ we also write $f$.
Thus a tree consists of a symbol of rank $m$ together
with an $m$-tuple of trees. 
We fix the sets of input variables $X=\{x_1,x_2,\dots\}$ and
formal parameters $Y=\{y_1,y_2,\dots\}$
where for
$m\in\Nat$, $X_m = \{x_1,\ldots,x_m\}$
and $Y_m = \{y_1,\dots,y_m\}$.

A (\emph{deterministic top-down}) \emph{tree automaton} ($\DTTA$ automaton for short)  is a
tuple $A=(P,\Sigma,p_0,\rho)$ where $P$ is a finite set of states,
$\Sigma$ a ranked alphabet, $p_0\in P$ the initial state,
and $\rho$ the transition function. 
For every $f\in\Sigma^{(m)}$ and $p\in P$,
$\rho(p,f)$ is undefined or is in $P^m$.
The transition function allows to define for each $p\in P$ the set ${\sf dom}(p)\subseteq \T_\Sigma$ by
letting $f\,\bt\in{\sf dom}(p)$ for $f\in\Sigma^{(m)}$, $m\geq 0$, 
and $\bt\in\T_\Sigma^m$ iff 
$\rho(p,f)=\bp$ and $\bt_i\in{\sf dom}(\bp_i)$ for $i\in[m]$. 
%
%
The language $\L(A)$ of $A$ is given by $\L(A) = {\sf dom}(p_0)$.
The \emph{size} $|A|$ of $A$ is defined as
$|P|+|\Sigma|+|\rho|$ where $|\rho|=\sum_{m\geq 0}|\rho^{-1}(P^m)|\cdot(m+1)$.


Let $l\geq 0$.
A \emph{deterministic macro tree-to-string transducer} (\emph{with $l$ parameters})
($y\DMTT$ transducer for short) 
is a tuple $M=(Q,\Sigma,\Delta,q_0,\delta)$, where
$Q$ is a ranked alphabet of states all of rank $l+1$,
$\Sigma$ is a ranked alphabet of input symbols,
$\Delta$ is an alphabet of output symbols,
$q_0\in Q$ is the initial state, and
$\delta$ is the transition function.
For every $q\in Q$, $m\geq 0$, and $f\in\Sigma^{(m)}$, 
$\delta(q,f)$ is either undefined or is in $R$, where $R$ is the smallest set such that
$\epsilon\in R$ and if $T,T_1,\dots,T_l\in R$, then also
\begin{enumerate}
\item[(1)] $aT\in R$ for $a\in\Delta$,
\item[(2)] $y_jT\in R$ for $j\in[l]$, and
\item[(3)] $q'(x_i,T_1,\dots,T_l)\,T\in R$ for $q'\in Q$ and $i\in[m]$.
\end{enumerate}
Again, we represent the fact that $\delta(q,f) = T$ also by the rule:
\[
q(f(x_1,\ldots,x_m),y_1,\ldots,y_l)\to T.
\]
A state $q\in Q$ induces a partial function $\sem{q}_M$ from $\T_\Sigma$ 
to total functions $(\Delta^*)^l\to\Delta^*$ defined recursively as follows.
Let $t=f\,\bt$ with $f\in\Sigma^{(m)}$, $m\geq 0$, and $\bt\in\T_\Sigma^m$.
Here, we consider a \emph{call-by-value} (or \emph{inside-out}) mode of evaluation
for the arguments of states. 
Thus for $\bw\in(\Delta^*)^l$, 
$\sem{q}_M(t)(\bw)$ is defined whenever $\delta(q,f)=T$ for some $T$
and for each occurrence of a subtree $q'(x_i,T_1,\ldots,T_l)$ in $T$,
$\sem{q'}_M(\bt_i)$ is also defined. In this case, the output is obtained by evaluating 
$T$ in call-by-value order with $\bt_i$ taken for $x_i$ and $\bw_j$ for $y_j$.
In function applications (especially for higher-order) we often leave out 
parenthesis; e.g. we write $\sem{T}_M\,\bt\,\bw$ for $\sem{T}_M(\bt)(\bw)$.
We obtain,
\[
\sem{q}_M(f\,\bt)\,\bw = \sem{T}_M\,\bt\,\bw
\]
where the evaluation function $\sem{T}_M$ is defined as follows:
\[
\begin{array}{rll}
\sem{\epsilon}_M\, \bt\, \bw	&=&	\epsilon	\\
\sem{aT'}_M\, \bt\, \bw		&=&	a\;\sem{T'}_M\,\bt\,\bw	\\
\sem{y_jT'}_M\, \bt\, \bw	&=&	\bw_j\,\sem{T'}_M\,\bt\,\bw	\\
\sem{q'(x_i,T_1,\ldots,T_l) T'}_M\,\bt\,\bw
	&=&	
	\sem{q'}_M\;\bt_i\;
	(\sem{T_1}\,\bt\,\bw,\ldots,\sem{T_l}\,\bt\,\bw)\;\sem{T'}_M\,\bt\,\bw.
\end{array}
\]
The transducer $M$ realizes the (partial) translation $M:\T_\Sigma\to\Delta^*$ which,
for $t\in\T_\Sigma$, is defined as $M(t)=\sem{q_0}_M(t)(\epsilon,\ldots,\epsilon)$ 
if $\sem{q_0}_M(t)$ is defined and is undefined otherwise; the domain of this 
translation is denoted ${\sf dom}(M)$.
The $y\DMTT$ transducer $M$ is
a \emph{deterministic top-down tree-to-string transducer} 
($y\DT$ transducer for short) if $l=0$. 
The $y\DMTT$ transducer $M$ is
\emph{total} if $\delta(q,f)$ is defined 
for all $q\in Q$ and $f\in\Sigma$, and
$M$ is \emph{unary} if $|\Delta|=1$.
In the latter case, the output can also be represented by a number, namely, the length of the output.
Finally, a $y\DMTT$ transducer is \emph{self-nested}, if there is a right-hand side $T$ in $\delta$ 
so that $T$ contains an occurrence of a tree $q'(x_i,T_1,\ldots,T_l)$ where one of the trees
$T_j$ contains another tree $q''(x_i,T'_1,\ldots,T'_l)$ for the same $x_i$.
A $y\DMTT$ transducer is called \emph{non-self-nested}, if it is \emph{not} self-nested.

As for $\DTTA$ automata, 
we define the size $|M|$ of a $y\DT$ or $y\DMTT$ transducer $M$ as the sum of the sizes of the involved
alphabets, here $Q,\Sigma$ and $\Delta$, together with the size of the corresponding
transition function where the size of a transition $\delta(q,f) = T$ is one plus the sum of
numbers of occurrences of output symbols, parameters, and states in $T$. 
%

\medskip

\noindent
In the following three sections, we consider transducers with unary output alphabet $\Delta=\{d\}$ only.
In this case, we prefer to let the transducer 
produce the \emph{lengths} of the output directly. Then, the right-hand sides
$T$ may no longer contain symbols $d$, but constant numbers $c$ (representing $d^c$). 
Likewise, concatenation is replaced with addition. 
For convenience, we also allow multiplication with constants to compactly represent
repeated addition of the same subterm.
\noindent
\begin{example}\label{e:unary}
\rm
Consider the $y\DMTT$ transducer with set $Q=\{q_0,q\}$ of states and 
initial state $q_0$ and the following transition rules:
\[
\begin{array}{lcl}
q_0(f(x_1,x_2),y_1)&\to& q(x_1,q(x_2,d))\\
q(a(x_1),y_1)&\to& y_1 q(x_1,y_1)\\
q(e,y_1)&\to& \epsilon	\\
\end{array}
\]
where the output is considered as a string. This $y\DMTT$ transducer is non-self-nested and
realizes a translation $\tau$ which maps each input tree 
$f(a^n(e),a^m(e))$ to the string in $d^{n\cdot m}$.
As the output alphabet is unary, we prefer to represent
the output length by these rules:
\[
\begin{array}{lcl}
q_0(f(x_1,x_2),y_1)&\to& q(x_1,q(x_2,1))\\
q(a(x_1),y_1)&\to& y_1 + q(x_1,y_1)\\
q(e,y_1)&\to& 0.	\\
\end{array}
\]
A $\DTTA$ automaton accepting the domain of the given $y\DMTT$ transducer may use states from $\{p_0,p\}$
with initial state $p_0$
where
\[
\rho(p_0,f)	= (p,p)	\qquad
\rho(p,a)	= p\qquad
\rho(p,e)	=().
\]
\qed
\end{example}
Thus, right-hand sides $T$ now are constructed according to the grammar:
\[
\begin{array}{lcl}
T	&{::=}&	c\mid y_j\mid q(x_i,T_1,\ldots,T_l)	\mid  T_1+T_2	\mid c\cdot T'
\end{array}
\]
where the non-negative numbers $c$ may be taken from some fixed range $\{0,1,\ldots,h\}$.
The size $|T|$ then is defined as the size of $T$ as an expression, i.e.,
\[
\begin{array}[t]{lll}
|c|&=& |y_j| = 1	\\
|q(x_i,T_1,\ldots,T_l)| &=& 2 + |T_1|+\ldots+|T_l|	\\
\end{array}\qquad\qquad
\begin{array}[t]{lll}
|T_1+T_2|	&=& |T_1|+|T_2|	\\
|c\cdot T'|	&=& 2+|T'|.
\end{array}
\]
Thus, e.g., for $T = 2+3\cdot q(x_1,1,0)$, $|T| = 4 + |q(x_1,1,0)| = 4+4 = 8$.

\medskip

\subsection*{From Arbitrary to Binary Input Alphabets}

Here we state a technical lemma that allows to restrict
the rank of output symbols of our transducers to two.
Let $\Sigma$ be a ranked alphabet and $\bot$ 
a special symbol not in $\Sigma$. 
By $\bin(\Sigma)$ we denote the ranked alphabet 
$\{\bot^{(0)}\}\cup\{\sigma^{(2)}\mid\sigma\in\Sigma\}$. 
For sequences $s$ of trees over $\Sigma$ we define their
\emph{binary encoding} $\bin(s)$ as:
$\bin(s)=\bot$ if $s$ is the empty sequence, and
$\bin(s)=\sigma(\bin(t_1t_2\cdots t_m), \bin(s'))$
if $s=\sigma(t_1,\dots,t_m)s'$ with 
$\sigma\in\Sigma^{(m)}$, $m\geq 0$, $t_1,\dots,t_m\in\T_\Sigma$,
and $s'\in\T_\Sigma^*$.
Likewise for $S\subseteq\T_\Sigma$, $\bin(S) = \{\bin(s)\mid s\in S\}$.
Note that this encoding corresponds to the \emph{first-child-next-sibling} encoding
of unranked trees, here applied to ranked trees.

As an example, consider the tree $t = f(b,g(c),h(b,c))$. Then
the encoding $\bin(t)$ is given by:
\[
\bin(t) = f(b(\bot,g(c(\bot,\bot),h(b(\bot,c(\bot,\bot)) ,\bot)))
	,\bot)
\]

\begin{lemma}\label{l:bin}
Let $M=(Q,\Sigma,\Delta,q_0,\delta)$ be a $y\DMTT$ transducer
and let $m$ be the maximal rank of symbols in $\Sigma$. 
Then a $y\DMTT$ transducer $M'=(Q',\bin(\Sigma),\Delta,q'_0,\delta')$
together with a $\DTTA$ automaton $B$ 
can be constructed in time polynomial in $|M|$ such that 
\begin{enumerate}
\item[(1)] $\bin(\dom(M)) = \L(B)\cap\dom(M')$,
\item[(2)] $M'(\bin(t))=M(t))$ for all $t\in\dom(M)$,
\item[(3)] $|Q'|=m|Q|$,
\item[(4)] $M'$ is a total $y\DT$ if $M$ is.
\end{enumerate}
\end{lemma}
\begin{proof}
The $\DTTA$ automaton $B$ is meant to check whether a tree in
$\bin(\Sigma)$ is an encoding of a tree in $\bin(\T_\Sigma)$,
i.e., $\L(B)=\bin(\T_\Sigma)$. Such an automaton can be constructed 
as $B=(P,\bin(\Sigma),1,\rho)$ where $P=\{0\}\cup[m]$ and $\rho$ is
given by:
\[
\begin{array}{lll@{\quad}l}
\rho(j+1,f) 	&=& (k,j)	&\text{if}\;f\in\Sigma^{(k)}	\\
\rho(0,\bot)	&=& ()				\\
\end{array}
\]
For this $\DTTA$ automaton $B$,
$\L(B) =\bin(\T_\Sigma)$ holds.

The $y\DMTT$ transducer $M'$ is defined as follows.
$Q'=\{\angl{q,i}\mid q\in Q, m\geq 1, i\in[m]\}$ 
and $q'_0=\angl{q_0,1}$.
Let $q\in Q$ and $f\in\Sigma$ of arity $k\geq 0$.
If $\delta(q,f)$ is defined and equals $T$, then we
let $\delta'(\angl{q,1},f)=T'$ where $T'$ is
obtained from $T$ by replacing every occurrence of
$q'(x_i,T_1,\dots,T_l)$ with $\angl{q',i}(x_1, T'_1,\dots,T'_l)$ where
each $T'_i$ is obtained from $T_i$ in the same way.
Furthermore for every $2\leq i\leq k$, we define
$\delta'(\angl{q,i},f)=\angl{q,i-1}(x_2,y_1,\dots,y_l)$ and
finally for all $q'\in Q'$, $\delta'(q',\bot)=\epsilon$.

By construction, $M'$ is total  whenever $M$ is. Also, the
bounds to the number of states is obvious. For 
the correctness of the construction, we observe that whenever for $i\geq 1$,
the state $\angl{q,i}$ is called with the encoding of a list $t_1\ldots t_k$ with
$k\geq i$, then the same output is produced as by $M$ when applied to the input tree $t_i$,
i.e., 
\[
\sem{\angl{q,i}}_{M'}({\sf bin}(t_1\ldots t_k))(y_1,\ldots,y_l)
= \sem{q}_M(t_i)(y_1,\ldots,y_l)
\]
The proof is by induction on the structure of the sequence $t_1\ldots t_k$,
where for subterms $T$ of right-hand sides of $M$ and corresponding subterms $T'$
of right-hand sides of $M'$, 
\[
\begin{array}{lll}
\sem{T'}_{M'}({\sf bin}(t_1\ldots t_k), s') &=& \sem{T}_M(t_1,\ldots,t_k)
\end{array}
\]
holds for all $s'$.
From this, the first two assertions follow.
\end{proof}

\noindent
\begin{example}\rm
Consider the following rule:
\[
q(f(x_1,x_2,x_3),y_1) \to q(x_1,q(x_2,q(x_3,y_1)))
\]
for the ternary symbol $f\in\Sigma$. By the construction provided for Lemma~\ref{l:bin},
this rule is simulated by means of the rules:
\[
\begin{array}{lll}
q_1(f(x_1,x_2),y_1) &\to& q_1(x_1, q_2(x_1,q_3(x_1,y_1)))	\\
q_2(g(x_1,x_2),y_1) &\to& q_1(x_2,y_1)	\hfill(g\in\Sigma)\\
q_3(g(x_1,x_2),y_1) &\to& q_2(x_2,y_1)	\hfill(g\in\Sigma)\\
\end{array}
\]
where the state $q_1$ corresponds to the state $q$ and the states $q_2,q_3$ 
traverse the encoding of a list $t_1t_2t_3$ where $q$ is called for $t_2$ and $t_3$,
respectively.
\qed
\end{example}
\noindent
We remark that non-self-nestedness may not be preserved by our construction.


%
%
%

\newcommand{\Aff}{{\sf aff}}
\section{Unary Transducers without Parameters}\label{s:total}

\noindent
We first consider a single unary total $y\DT$ transducer 
and show that equivalence of two states relative to a $\DTTA$ automaton can
be decided in polynomial time. This result then is extended to decide equivalence of two not necessarily total
unary $y\DT$ transducers.
Let $M=(Q,\Sigma,\{d\},q_0,\delta)$ be a total unary $y\DT$ transducer, and 
$A=(P,\Sigma,p_0,\rho)$ a $\DTTA$ automaton.
Assume that $Q=[n]$ for some natural number~$n$.
Our goal is to decide for given $q,q'\in Q$ whether or not
$\sem{q}_M(t)=\sem{q'}_M(t)$ for all $t\in\L(A)$.
%
For every $t\in\T_\Sigma$ and $q\in Q$,
$\sem{q}_M(t)=d^r$ with $r\in\Nat$, i.e.,
$\sem{q}_M$ 
can be seen 
as a tree-to-integer translation mapping
$t$ to $r$; we denote $r$ by $\sem{t}_q$ 
and write 
$\sem{t}$ for the vector $(\sem{t}_{1},\dots,\sem{t}_{n})\in\Nat^n$, or, more generally, in ${\mathbb Q}^n$.
For a vector ${\bf v}\in{\mathbb Q}^n$ we again denote its $i$th component
by ${\bf v}_i$.
Then for $q\in Q$, $m\geq 0$, $f\in\Sigma^{(m)}$, and $\bt_1,\dots,\bt_m\in\T_\Sigma$, 
the output $\sem{f(\bt_1,\ldots,\bt_m)}_{q}\in{\mathbb Q}$ can be computed arithmetically by
\begin{eqnarray}
\sem{f(\bt_1,\ldots,\bt_m)}_{q} = \sem{\delta({q},f)}_M(\sem{\bt_1},\ldots,\sem{\bt_m}) 
\label{d:trans}
\end{eqnarray}
where for $T\in(\Delta\cup Q(X_m))^*$
and a vector ${\bf x}=({\bf x}_1,\ldots,{\bf x}_m)$ of vectors ${\bf x}_i\in{\mathbbm Q}^n$ the number 
$\sem{T}_M\;{\bf x}$ is given by:
\[
\begin{array}{lll}
\sem{c}_M\;{\bf x}&=&c	\\
\sem{j(x_i)}_M\;{\bf x}    &=&	{\bf x}_{ij}	\\
\sem{T'_1+T'_2}_M\;{\bf x}&=&\sem{T'_1}_M\;{\bf x}+
						      \sem{T'_2}_M\;{\bf x}	\\
\sem{c\cdot T'}_M\;{\bf x}&=&c\cdot\sem{T'}_M\;{\bf x}.	\\
\end{array}
\]
By structural induction on $T$, we conclude that 
\[
\sem{T}_M\;{\bf x} = b_0 + \sum_{i=1}^m\sum_{j=1}^n b_{ij}\cdot {\bf x}_{ij}
\]
for suitable numbers $b_0,b_{ij}\in{\mathbb Q}$.
Thus, $\sem{T}_M$ and hence also $\sem{\delta(q,f)}_M$ constitutes 
a \emph{multi-affine} mapping from $({\mathbbm Q}^n)^m$ to ${\mathbbm Q}$. 
Technically, a multi-affine mapping $H$ is affine in each argument.
This means that the transformation $H'$ corresponding to the $k$th argument and
fixed ${\bf x}_1,\ldots,{\bf x}_{k-1}, {\bf x}_{k+1},\ldots,{\bf x}_m$, which is
defined by:
\[
H'({\bf x}') = H({\bf x}_1,\ldots,{\bf x}_{k-1}, {\bf x}',
                         {\bf x}_{k+1},\ldots,{\bf x}_m)
\]
is  affine, i.e.,
\[
H'({\bf y}_0 + \sum_{r=1}^n\lambda_r({\bf y}_r-{\bf y}_0)) =
H'({\bf y}_0) + \sum_{r=1}^n\lambda_r (H'({\bf y}_r)-H'({\bf y}_0))
\]
holds
for vectors ${\bf y}_0,\ldots,{\bf y}_n\in{\mathbbm Q}^n$ and 
$\lambda_1,\ldots,\lambda_n\in{\mathbbm Q}$.
Accordingly, we define the (output) semantics of 
$f\in\Sigma$
of arity $m$ as the function $\sem{f}:({\mathbbm Q}^n)^m\to{\mathbbm Q}^n$ by:
\begin{eqnarray}
\sem{f} \;{\bf x} = 
	(\sem{\delta(1,f)}_M\;{\bf x},\ldots,\sem{\delta(n,f)}_M\;{\bf x})
\end{eqnarray}
which again is multi-affine.

 %

\begin{figure}[hbt]
\[
\begin{array}{l}
{\bf forall}\;(p\in P)\;B_p\,{:=}\,\emptyset;	\\
{\bf repeat}\\
\qquad{\sf done}\,{:=}\,{\bf true};	\\
\qquad{\bf forall}\; \;(p,p_1,\ldots,p_m\in P, f\in\Sigma\;\text{with}\;\rho(p,f) = (p_1,\ldots, p_m))\;\\
\qquad\qquad{\bf forall}\;(({\bf v}_1,\ldots,{\bf v}_m)\in B_{p_1}\times\ldots\times B_{p_m})\;	\\
\qquad\qquad\qquad{\bf v}\,{:=}\,\sem{f}({\bf v}_1,\ldots,{\bf v}_m);	\\
\qquad\qquad\qquad{\bf if}\;{\bf v}\not\in\Aff(B_p)\;	\\
\qquad\qquad\qquad\qquad B_p\,{:=}\,B_p\cup\{{\bf v}\};	\\
\qquad\qquad\qquad\qquad {\sf done}\,{:=}\,{\bf false};	\\
{\bf until}\;({\sf done}={\bf true});
\end{array}
\]
\caption{Computing bases for the closures $\Aff(\{\sem{t}\mid t\in\dom(p)\}),p\in P$.}
\label{alg:aff}
\end{figure}

\begin{theorem}\label{t:total}
Let $\Sigma$ be a fixed ranked alphabet, and $A$ a $\DTTA$ automaton over $\Sigma$.
Let $M$ a total unary $y\DT$ transducer with input alphabet $\Sigma$, and $q,q'$ states of $M$.  
It is decidable in polynomial time whether or not $\sem{q}_M(t)=\sem{q'}_M(t)$
for all $t\in\L(A)$.
\end{theorem}
\begin{proof}
By repeated application of the transformations $\sem{f}, f\in \Sigma$, every tree $t\in\T_\Sigma$ gives rise to a vector 
$\sem{t} \in {\mathbbm Q}^n$.
%
For a set $S\subseteq\T_\Sigma$, let $\sem{S}=\{\sem{t}\mid t\in S\}$.
Then two states $q,q'$ are equivalent relative to $S\subseteq\T_\Sigma$ iff 
$H_{qq'}({\bf v}) = 0$ for all ${\bf v}= ({\bf v}_1,\ldots,{\bf v}_n)\in \sem{S}$,
where the function $H_{qq'}$ is given by $H_{qq'}({\bf v}) = {\bf v}_{q}-{\bf v}_{q'}$.
The set of vectors in $\sem{\L(A)}$
can be characterized by means of a constraint system.
Consider the collection of
sets $V_p,p\in P$, which are the least sets with
\begin{eqnarray}
V_p\supseteq \sem{f}(V_{p_1},\ldots,V_{p_m})
\label{eq:constraints}
\end{eqnarray}
whenever $\rho(p,f) =  (p_1,\ldots,p_m)$ holds.
Then $\{\sem{t}\mid t\in\L(A)\}$ is precisely given by the set $V_{p_0}$.

For a set $V\subseteq{\mathbbm Q}^n$ of $n$-dimensional vectors, let
$\Aff(V)$ denote the \emph{affine closure} of $V$.
This set is obtained from $V$ by adding all affine combinations of elements in $V$: 
\[
\Aff(V)=\{{\bf s}_0+\sum_{j=1}^r\lambda_j\cdot ({\bf s}_j-{\bf s}_0)\mid 
r\geq 0,{\bf s}_0,\dots, {\bf s}_r\in V,\lambda_1,\ldots,\lambda_r\in{\mathbbm Q}\}.
\]
Every set $V\subseteq{\mathbb Q}^n$ has a subset $B\subseteq V$ of cardinality at most $n+1$ such
that the affine closures of $B$ and $V$ coincide. A set $B$ with this property of minimal
cardinality is also called \emph{affine basis} of $\Aff(V)$.
For an affine function $H:{\mathbb Q}^n\to{\mathbb Q}$ such as $H_{qq'}$ and every subset 
$V\subseteq {\mathbb Q}^n$, the following three statements are equivalent:
\begin{enumerate}
\item	$H({\bf v})=0$ for all ${\bf v}\in V$; 
\item	$H({\bf v})=0$ for all ${\bf v}\in \Aff(V)$;
\item	$H({\bf v})=0$ for all ${\bf v}\in B$ if $B$ is any subset of $V$ with $\Aff(B) = \Aff(V)$.
\end{enumerate}
Instead of verifying that $H({\bf v})=0$ holds for all elements $\bf v$ of $V_{p_0}$,
it suffices to test $H({\bf v})=0$ for all elements $\bf v$ of an affine basis $B\subseteq V_{p_0}$.
%
Accordingly, we are done if we succeed in computing an affine basis of the set $\Aff(V_{p_0})$.
It is unclear, though, how the least solution $V_p,p\in P$, of the constraint system \eqref{eq:constraints}
can be computed. 
Instead of computing this least solution,
we propose to consider the least solution of the constraint system \eqref{eq:constraints}, 
not over \emph{arbitrary} subsets but over \emph{affine} subsets of ${\mathbb Q}^n$ only.
%
Like the set ${\cal P}({\mathbb Q}^n)$ of all subsets of ${\mathbb Q}$ (ordered by subset inclusion),
the set ${\cal A}({\mathbb Q}^n)$ of all affine subsets of ${\mathbb Q}$ (still ordered by subset inclusion) forms a complete lattice, 
but where the least upper bound operation is not given by set union. 
Instead, for a family $\cal B$ of affine sets, 
the least affine set containing all $B\in{\cal B}$ is given by:
\[
\mbox{$\bigsqcup{\cal B} = \Aff(\bigcup{\cal B})$}.
\]
We remark that affine mappings commute with 
least upper bounds,
i.e., for every affine mapping $F:{\mathbb Q}^n\to{\mathbb Q}^n$,
\[
\begin{array}{lll}
F(\bigsqcup{\cal B}) &=& F(\Aff(\bigcup{\cal B})) = \Aff(F(\bigcup{\cal B})) \\
&=& \Aff(\{F({\bf v)})\mid {\bf v}\in\bigcup{\cal B}\})	\\
&=& \bigsqcup\{F(B)\mid B\in{\cal B}\}.
\end{array}
\]
Let $V_p,p\in P$, and $V^\sharp_p,p\in P,$ denote the least solutions of \eqref{eq:constraints} 
w.r.t.\ the complete lattices 
${\cal P}({\mathbb Q}^n)$ and ${\cal A}({\mathbb Q}^n)$, respectively.
Since for each $f\in\Sigma$, $\sem{f}$ is affine in each of its arguments,
it follows by the 
transfer lemma of \cite{countable86} (see also \cite{fix95}), that
\[
\Aff(V_p) = V^\sharp_p\qquad(p\in P).
\]
The complete lattice 
${\cal A}({\mathbb Q}^n)$, on the other hand, 
satisfies the \emph{ascending chain} condition. 
This means that every increasing sequence of affine subsets is ultimately stable. 
Therefore, the least solution $V^\sharp_p,p\in P,$ of the constraint system \eqref{eq:constraints} 
over ${\cal A}({\mathbb Q}^n)$ can be effectively computed by fixpoint iteration.
One such fixpoint iteration algorithm is presented in Figure~\ref{alg:aff}.
Each occurring affine subset of ${\mathbb Q}^n$ is represented by a basis. 
For the resulting basis $B_{p_0}$ of the affine subset $V^\sharp_{p_0} = \Aff(V_{p_0})$
we finally may check whether or not $H_{qq'}({\bf v}) = 0$ for all ${\bf v}\in B_{p_0}$,
which completes the procedure.

The algorithm of Figure~\ref{alg:aff} 
starts with empty sets $B_p$ for all $p\in P$. 
Then it repeatedly performs one round through all transitions $\rho(p,f) =(p_1,\ldots,p_m)$ of $A$
while the flag ${\sf done}$ is \textbf{false}.
For each transition
$\rho(p,f)=(p_1,\ldots p_m)$, the transformation $\sem{f}$ is applied to every 
$m$-tuple ${\bf v}=({\bf v}_1,\ldots,{\bf v}_m)$ with ${\bf v}_i\in B_{p_i}$. 
The resulting vectors then are added to $B_p$ --- 
whenever they are not yet contained in the affine closure $\Aff(B_p)$ of $B_p$. 
The iteration terminates when during a full round of the \emph{repeat-until} loop, 
no further element has been added to any of the $B_p$.

In the following, we assume a uniform cost measure where arithmetic operations are counted as 1.
Thus, evaluating a right-hand side $\delta(q,f)$ takes time at most proportional to 
the number of symbols occurring in $\delta(q,f)$. 
Concerning the complexity of the algorithm, we note:
\begin{ylist}{\textbullet}
\item
The algorithm performs at most $h\cdot (n+1)$ rounds on the \emph{repeat-until} loop
($h$ and $n$ are the number of states of $A$ and $M$, respectively);
\item
In each round of the \emph{repeat-until} loop, for each transition of~$A$, at most $(n+1)^m$ tuples are considered
($m$ is the maximal arity of an input symbol);

\item
for each encountered vector, time $O(n^3)$ is sufficient to check whether the vector is contained in
the affine closure of the current set $B_p$ (see, e.g., chapter 28.1 of \cite{Cormen}).
\end{ylist}

\noindent
Accordingly, a full round of the \emph{repeat-until} loop
can be executed in time $O(|A|\cdot|M|\cdot n^m)$ --- giving us an upper complexity bound 
$O(|A|\cdot|M|\cdot h n^{m+4})$ for the algorithm
%
where $m$ can be chosen as 2,
according to Lemma~\ref{l:bin}.
%
\end{proof}

\noindent
The base algorithm as presented in the proof of Theorem~\ref{t:total},
can be further improved as follows:
\begin{ylist}{\textbullet}
\item
We replace the Round-robin iteration by a worklist iteration which re-schedules the evaluation
of a transition of $A$
for a state $p'\in P$ and an input symbol $f$
only if $B_{p_i}$ for one of the states $p_i$ in $\rho(p',f)$ 
has been updated. 
\item
We keep track of the set of tuples which have already been processed for a given pair $(p',f)$, $f$ an input symbol 
and $p'$ state of $A$.
This implies that throughout the whole fixpoint iteration, for each such pair $(p',f)$,
inclusion in the affine closure must only be checked for $(n+1)^m$ tuples.
\item
For a non-empty affine basis $B$, we can maintain a single element $v'\in B$, together
with a basis of the linear space $L_B$ corresponding to $B$, spanned by the vectors
$({v}-v'), {v}\in B\backslash\{v'\}$.
By maintaining a basis of $L_B$ in \emph{Echelon} form, membership in $\Aff(B)$ can be tested in
time $O(n^2)$. 
\end{ylist}
Applying these three optimizations, the overall complexity comes down to $O(|A|\cdot |M|\cdot n^{m+2})$.

So far, we have compared the output behavior of two states  of a unary total $y\DT$ transducer $M$ 
relative to some $\DTTA$ automaton only. Our decision procedure for equivalence, however, readily 
extends to arbitrary unary $y\DT$ transducers.
Note that the exponential upper bound of Theorem~\ref{t:partial} is sharp, 
since non-emptiness for unary $y\DT$ transducers is already
{\sc Exptime}-complete (see Theorem~9 of~\cite{DBLP:journals/corr/Maneth14}).

\begin{theorem}\label{t:partial}
Equivalence for (possibly partial) unary $y\DT$ transducers
can be decided in deterministic exponential time.
If the transducers are linear, then the 
algorithm runs in polynomial time.
\end{theorem}
\begin{proof}
First, we, w.l.o.g., may assume that we are given two states $q_0,q'_0$ of a single $y\DT$ transducer,
and the task is to decide whether the partial mappings $\sem{q_0}_M$ and $\sem{q_0'}_M$ coincide, i.e., whether
(1) $\sem{q_0}_M(t)$ is defined iff $\sem{q_0'}_M(t)$ is defined, and (2) $\sem{q_0}_M(t) = \sem{q_0'}_M(t)$ whenever
both are defined.
In order to decide the former task, we construct $\DTTA$ automata $A,A'$ where the languages 
of $A$ and $A'$ are precisely given by the domains of the translations
$\sem{q_0}_M$ and $\sem{q'_0}_M$, respectively.
 
The set of states and transitions of $A$ can be determined as the smallest subset $P$ of sets $Q'\subseteq [n]$
together with the partial function $\rho$ as follows.
First, $\{q_0\}\in P$ which also serves as the initial state of $A$. 
Then for every element $Q'\in P$ and every input symbol $f\in\Sigma$ of some arity $m$,
where $\delta(q,f)$ is defined for each $q\in Q'$,
every set $Q'_i$ is contained in $P$ for $i=1,\ldots,m$, where $Q'_i$ is the set of all states $q'\in[n]$ such 
that $q'(x_i)$ occurs in the right-hand side $\delta(q,f)$ for some $q\in Q'$. In this case then $\rho$ has the
transition $\rho(Q',f) = Q'_1\ldots Q'_m$.
The $\DTTA$ automaton $A'$ is obtained by starting with
the initial state $\{q_0'\}$ instead of $\{q_0\}$, and subsequently proceeding analogously to the construction of $A$. 
Assume that the number of states of $A$ and $A'$ are $q$ and $q'$, respectively. 
Then property (1) is satisfied iff $\L(A)=\L(A')$. This can be verified in time polynomial in the sizes of $A$ and $A'$.
Now assume that $\L(A)=\L(A')$.
%
Then we construct a total $y\DT$ transducer $M'$ from $M$ by adding to the transition function of $M$
a transition $q(f(x_1,\ldots,x_m)) \to \epsilon$ for every 
state $q$ and input symbol $f$ --- where $M$ does not yet provide a transition.
By construction, $\sem{q}_{M'}(t) = \sem{q}_M(t)$ whenever $\sem{q}_M(t)$ is defined.
Therefore for every $t\in\L(A)$, $\sem{q_0}_{M'}(t) = \sem{q'_0}_{M'}(t)$ iff $\sem{q_0}_{M}(t) = \sem{q'_0}_{M}(t)$.
Using the algorithm of Theorem~\ref{t:total}, the latter can be decided in time polynomial 
in the sizes of $A$ and $M'$.

The size of the $\DTTA$ automaton $A$ characterizing the domain of the $y\DT$ transducer $M$ 
is at most exponential in the size of $M$. 
In case, however, that $M$ is linear, the size of the corresponding automaton $A$ is at most linear in
the size of $M$. From that, the complexity bounds of the theorem follow.
\end{proof}

\noindent
Theorems~\ref{t:total} and~\ref{t:partial} can be applied to decide 
\emph{Abelian} equivalence of $y\DT$ transducers with arbitrary output alphabet.
\emph{Abelian} equivalence of two deterministic tree-to-string transducers means that the outputs for every input tree
coincide up to the ordering of output symbols. 
%
%

\section{Non-Self-Nested Unary Transducers with Parameters}\label{s:unnested}\label{s:nonnested}

We consider unary \emph{non-self-nested} $y\DMTT$ transducers and show
that their equivalence problem can be solved in co-randomized polynomial
time. This implies equivalence with the same complexity for (arbitrary) linear $y\DT$ transducers.

Deterministic macro tree-to-string transducers ($y\DMTT$ transducers) combine $y\DT$ transducers 
with the
nesting present in macro grammars. 
Each state of a $y\DMTT$ transducer takes a fixed number of \emph{parameters} (of type output tree).
Recall that a $y\DMTT$ transducer is non-self-nested 
if whenever $q'(x_j,\ldots)$ occurs nested in $q(x_i,\ldots)$ implies that $i\neq j$.
%
%
Note that non-self-nested $y\DMTT$ transducers are strictly more powerful than
$y\DT$ transducers
as shown in the following lemma.

\begin{lemma}\label{l:nonnested}
The translation of Example~\ref{e:unary}, which is realized by a non-self-nested 
unary $y\DMTT$ transducer,  
cannot be realized by any $y\DT$ transducer.
\end{lemma}

\begin{proof}
For convenience, we prove a slightly stronger result, namely, that this translation
also cannot be realized by any $y\DT$ transducer 
even if it is equipped with \emph{regular look-ahead} (a $y\DTR$ transducer).
Assume for a contradiction, that a given $y\DTR$ transducer $N$ realizes the translation of $M$
where $N$ has a finite set $Q$ of states and 
uses the finite bottom-up automaton $B$ for providing 
look-ahead information about the input. 
Let 
$n_1\neq n_2$ so that $a^{n_1}(e)$ and $a^{n_2}(e)$ correspond to 
the same look-ahead state of $B$.
Then for $i=1,2$ and $j=1,2$,
\[
\begin{array}{l}
N(f(a^{n_i}(e),a^{n_j}(e))) = 
c +\sum_{q\in Q} c_{q} \sem{q}_N(a^{n_i}(e))
			      +\sum_{q\in Q} c'_{q} \sem{q}_N(a^{n_j}(e))
\end{array}
\]
for suitable numbers $c, c_{q},c'_{q}$ (independent of $i,j$). 
For $j=1,2$, consider the \emph{difference} in the outputs:
\[
\begin{array}{lll}
\Delta_j	&=& N(f(a^{n_1}(e),a^{n_j}(e))) - N(f(a^{n_2}(e),a^{n_j}(e)))		\\
		&=&	\sum_{q\in Q} c_{q} (\sem{q}_N(a^{n_1}(e))-\sem{q}_N(a^{n_2}(e)))
\end{array}
\]
and observe that it is independent of $j$. According to our assumption, $N$ realizes the
translation of $M$. Therefore, 
\[
\begin{array}{lll}
0	&=& 	\Delta_1 - \Delta_2	\\
	&=&	(n_1-n_2)\cdot n_1 - (n_1-n_2)\cdot n_2	\\
	&=&	(n_1-n_2)\cdot (n_1-n_2)	\\
	&\neq&	0
\end{array}
\]
--- a contradiction.
Hence the translation of $M$ cannot be realized by any $y\DTR$ transducer.
\end{proof}

As in the case for unary $y\DT$ transducers, 
we first consider \emph{total} unary $y\DMTT$ transducers only, but relative to
a $\DTTA$ automaton $A$. 
Assume that a unary $y\DMTT$ transducer $M$ is given by $M=([n],\Sigma,\{d\},i_0,\delta)$.
Recall that we assume that all states have exactly $l+1$ parameters where the first one is the
input tree and the remaining $l$ parameters accumulate output strings, i.e., numbers.
%
The output for a state $q$ and an $m$-ary input symbol $f\in\Sigma$, 
then is given by:
\begin{eqnarray}
\sem{f(\bt_1,\ldots,\bt_m)}_{q} {\bf y} = \sem{T}_M(\sem{\bt_1}\;{\bf y},\ldots,\sem{\bt_m}\;{\bf y})
\label{d:trans-1}
\end{eqnarray}
when $q(f(x_1,\ldots,x_m),y_1,\ldots,y_l)\to T$ is a rule of $M$, and
${\bf y}$ is a vector of parameters in ${\mathbb Q}^l$. Here,
$\sem{T'}\;{\bf x}\;{\bf y}$ for a vector 
${\bf x}=({\bf x}_1,\ldots,{\bf x}_m)$ of 
vectors ${\bf x}_i\in({\mathbbm Q}^{l}\to {\mathbbm Q})^n$ 
is defined by:
\[
\begin{array}{lll}
\sem{c}_M\;{\bf x}\;{\bf y}&=&c	\\
\sem{y_k}_M\;{\bf x}\;{\bf y} &=&{\bf y}_k\\
\sem{c\cdot T'}_M\;{\bf x}\;{\bf y}&=&c\cdot \sem{T'}\;{\bf x}\;{\bf y}	\\
\sem{T'_1+T'_2}_M\;{\bf x}\;{\bf y}&=&	
\sem{T'_1}_M\;{\bf x}\;{\bf y}+\sem{T'_2}_M\;{\bf x}\;{\bf y}\\
\sem{j(x_i,T'_1,\ldots,T'_l)}\;{\bf x}\;{\bf y} &=& 	
	{\bf x}_{ij}(\sem{T'_1}_M\;{\bf x}\;{\bf y},\ldots,\sem{T'_l}_M\;{\bf x}\;{\bf y}).
\end{array}
\]
%
%
By structural induction, we verify that for all input trees $t\in\T_\Sigma$ and all
states $q$ of the $y\DMTT$ transducer $M$,
$\sem{t}_q$ is an \emph{affine function} ${\mathbb Q}^l\to{\mathbb Q}$,
i.e., $\sem{t}_{q}\;{\bf y} = {\bf v}_{q0}+{\bf v}_{q1}{\bf y}_1+\ldots+{\bf v}_{ql}{\bf y}_l$ for suitable
${\bf v}_{qj}\in{\mathbb Q}$. Accordingly, $\sem{t}$ can be represented 
as the two-dimensional matrix ${\bf v}=({\bf v}_{qj})\in{\mathbb Q}^{n\times(l+1)}$.

Now assume that the arguments ${\bf x}_i,i=1,\ldots,m,$ 
are all vectors of affine functions ${\mathbb Q}^l\to{\mathbb Q}$,
and let ${\bf x}$ denote the triply indexed set $({\bf x}_{ijk})$ of coefficients in ${\mathbb Q}$
($i=1,\ldots,m$, $j=1,\ldots,n$ and $k=0,\ldots,l$) representing these functions.
Then
\begin{eqnarray}
\sem{f}_{q}\;{\bf x}\;{\bf y} &=&
	{\bf r}^{(f)}_{q0}+ {\bf r}^{(f)}_{q1}\cdot{\bf y}_1+\ldots+{\bf r}^{(f)}_{ql}\cdot{\bf y}_l
\label{eq:pol}
\end{eqnarray}
where ${\bf r}^{(f)}_{qk}$ is a polynomial
over the variables ${\bf x}$. Thus,
$\sem{f}$ can be represented by the matrix 
${\bf r}^{(f)}=({\bf r}^{(f)}_{jk})\in{\mathbb Q}[{\bf x}]^{n\times(l+1)}$.
\begin{example}\label{e:unary-1}\rm
Consider the $y\DMTT$ transducer $M$ from Example \ref{e:unary} which we extend to a total 
$y\DMTT$ transducer by adding the rules 
\[
q_0(a(e),y_1) \to 0\quad q_0(e,y_1)\to 0\quad q(f(x_1,x_2),y_1) \to 0
\]
Then we obtain:
\[
\begin{array}{lll}
\sem{f}_{q_0} ({\bf x}_1,{\bf x}_2) ({\bf y}_1)&=&
	{\bf x}_{1q0} + {\bf x}_{1q1}\cdot({\bf x}_{2q0} + {\bf x}_{2q1} \cdot 1)	\\
	&=&
	{\bf x}_{1q0} + {\bf x}_{1q1}\cdot{\bf x}_{2q0} + {\bf x}_{1q1}\cdot{\bf x}_{2q1}\\[1ex]
\sem{a}_q({\bf x}_1)({\bf y}_1)
	&=&
	{\bf y}_1 + {\bf x}_{1q0} + {\bf x}_{1q1}\cdot{\bf y}_1\\
	&=&
	{\bf x}_{1q0} + (1+ {\bf x}_{1q1})\cdot{\bf y}_1	\\[1ex]
\sem{e}_q()({\bf y}_1)
	&=&
	0.
\end{array}
\]
\qed
\end{example}	
\noindent
In this section, we first examine 
the case that the $y\DMTT$ transducer $M$ is non-self-nested (such as the $y\DMTT$ transducer from Example~\ref{e:unary}).
Then each polynomial ${\bf r}^{(f)}_{jk}$ in \eqref{eq:pol} is a sum of products:
\[
a\cdot {\bf x}_{i_1j_1k_1}\cdot\ldots\cdot{\bf x}_{i_{s}j_{s}k_{s}}
\]
where the $i_1,\ldots,i_s$ are pairwise distinct, i.e., each argument ${\bf x}_i$ 
contributes at most one factor to each product.
We conclude that
the transformation $\sem{f}$ is \emph{multi-affine}.
This means that
the mapping $\sem{f}$ when applied to an $m$tuple of values in ${\mathbb Q}^{n\times(l+1)}$
(i.e., vectors of \emph{affine} functions)
is an \emph{affine function} of each of the ${\bf x}_i$ and ${\bf y}$,
when the other arguments are kept constant.
Thus, $\sem{f}$ commutes with affine combinations in any of the arguments
${\bf x}_i$ and,
for each sequence ${\bf x}_1,\ldots,{\bf x}_m$ of matrices in ${\mathbb Q}^{n\times(l+1)}$,
again results in an affine function of ${\bf y}$.
%

As in the case of $y\DT$ transducers, we can construct a constraint system analogously to the system of 
constraints~\eqref{eq:constraints} whose unknowns 
are indexed with the states from the automaton $A$ ---
only that now each unknown $V_p$ receives a set of values in ${\mathbb Q}^{n\times(l+1)}$
(vectors of affine transformations)
instead of values in ${\mathbb Q}^n$
(plain vectors). 
This constraint system has a least solution where the value for $V_p$ is the set
of all affine transformations $\sem{t}, t\in\dom(p)$.

The two states $q_0,q'_0$ are equivalent with empty parameters relative to $A$
iff $H(\sem{t}) = 0$ for all $t\in\L(A)$
where 
$H({\bf v}) = {\bf v}_{q_00}-{\bf v}_{q'_00}$ for ${\bf v}=({\bf v}_{qk})\in{\mathbb Q}^{n\times(l+1)}$
(recall that for the affine function ${\bf v}_q=({\bf v}_{q0},\ldots,{\bf v}_{ql})$,
${\bf v}_q(0,\ldots,0) = {\bf v}_{q0}$).
As in the last section, the function $H$ for testing equivalence of states, is affine.

For a set $V\subseteq{\mathbbm Q}^{n\times(l+1)}$ of matrices, let
$\Aff(V)$ denote the \emph{affine closure} of $V$. This closure is defined analogously as for vectors.
Only note that now an affine basis of the affine closure of $V$ may have up to $n\cdot(l+1) + 1$ elements
(compared to $n+1$ in the last section).
Now let $H$ denote any affine function $H:{\mathbbm Q}^{n\times(l+1)}\to {\mathbbm Q}$.
Analogously to the last section, for every set $V\subseteq{\mathbbm Q}^{n\times(l+1)}$,
$H(v)=0$ holds for all $v\in V$ iff $H(v)=0$ holds for all $v$ in a basis of $\Aff(V)$.
We conclude that it suffices to determine for each state $p'$ of $A$, an \emph{affine basis} $B_{p'}$ of the set $V_{p'}$
and then verify that $H({\bf v}) = 0$ for all ${\bf v}\in B_{p_0}$ if $p_0$ is the
initial state of $A$.
With a similar algorithm as in the last section this is possible using a polynomial number
of arithmetic operations only --- given that the maximal arity of input symbols is bounded. 
Therefore, we obtain:

\begin{theorem}\label{t:nonnested}
Assume that $M$ is a non-self-nested total unary $y\DMTT$ transducer and $A$ is a $\DTTA$ automaton.
Then for every pair $q,q'$ of states of $M$, it is decidable whether $q$ and $q'$ are equivalent relative to $A$.
%
If the arity of input symbols is bounded by a constant,
the algorithm requires only a polynomial number of arithmetic operations. 
\end{theorem}


\noindent
In case of non-self-nested $y\DMTT$ transducers and multi-affine functions, the \emph{lengths} of occurring numbers, however,
can no longer be ignored. 
In order to calculate an upper bound to the occurring numbers, we first note
that for each state ${p'}$ of $A$, the basis of $\Aff(V_{p'})$ as calculated by our algorithm, is of the
form $\sem{t}$ for a tree  in $\L(A)$ of depth at most 
$((l+1)\cdot n+1)\cdot h$ if $n,l$ and $h$ are the numbers of states
and parameters of $M$, and the number of states of $A$, respectively. 
Concerning the lengths of occurring numbers, we prove:

\begin{lemma}\label{l:length}\label{l:count}
Assume that $M$ is a non-self-nested unary $y\DMTT$ transducer $M$ where the ranks of input symbols are bounded
by $m$, and the constants occurring in right-hand sides of rules are bounded by $h$. Then
\[
\sem{q}_M(t)({\bf y}_1,\ldots,{\bf y}_l)\leq
(h+1)^{(|M|\cdot(m+1))^N}\cdot b
\]
if $N$ is the depth of $t$ 
and $b$ is the maximum
of the argument numbers ${\bf y}_1,\ldots,{\bf y}_l$.
\end{lemma}
\begin{proof}
The proof is by induction on the depth of $t$. 
Thus, assume that $t = f\,{\bf t}$ with ${\bf t} = ({\bf t}_1,\ldots,{\bf t}_m)$, $m\geq 0$, 
and assume that the induction
hypothesis holds for the ${\bf t}_i$. Let $q(f(x_1,\ldots,x_m),y_1,\ldots,y_l)\to T$ be a
rule of $M$. Then for ${\bf y} = ({\bf y}_1,\ldots,{\bf y}_l)$,
\[
\begin{array}{lll}
\sem{q}_M(t)\,{\bf y} &=& \sem{T}_M\,{\bf t}\,{\bf y}
\end{array}
\]
For $T$ let $a(T)$ denote the nesting depth of calls $q(x_i,\ldots)$. Note that 
$a(T)\leq m$ since $M$ is assumed to be non-self-nested.
Since $|T|\leq |M|$, and the depth of each ${\bf t}_i$ is less than the depth of $t$,
the assertion follows from the following claim:
\begin{eqnarray*}
(\sem{T}_M\,{\bf t}\;{\bf y}) &\leq& (h+1)^{|T|\cdot (a(T)+1)\cdot (|M|\cdot(m+1))^{N-1}}\cdot b
\end{eqnarray*}
if $N$ is the maximal depth of a tree ${\bf t}_i$.
The proof of this claim is again by induction, but now 
on the structure of right-hand side $T$.

If $T$ is a constant or equals ${\bf y}_j$ for some $j$, the claim obviously holds.
In case $T$ equals a sum $T_1 + T_2$ or a scalar product $c\cdot T'$, the claim also follows
easily from the inductive hypothesis. It remains to consider the case where $T= q'(x_i,T_1,\ldots, T_l)$.
By inductive hypothesis for the $T_i$, we find that for every $i$, 
\[
\sem{T_i}_M\,{\bf t}\,{\bf y} \leq (h+1)^{|T_i|\cdot a(T)\cdot(|M|\cdot(m+1))^{N-1}}\cdot b
\]
since the nesting depth of each $T_i$ is at most $a(T)-1$.
Therefore,
\[
{\small
\begin{array}{lll}
\multicolumn{3}{l}{\sem{q'(x_i,T_1,\ldots, T_l)}_M\,{\bf t}\,{\bf y}) }\\
\phantom{\;}
	&=& \sem{q'}_M\,({\bf t}_i)\,(\sem{T_1}_M\,{\bf t}\,{\bf y},\ldots,\sem{T_l}_M\,{\bf t}\,{\bf y})\\
	&\leq&(h+1)^{(|M|\cdot(m+1))^{N-1}}\cdot (h+1)^{|T|\cdot a(M)\cdot (|M|\cdot(m+1))^{N-1}}\cdot b \\
	&\leq&(h+1)^{(|M|\cdot(m+1))^{N-1}+|T|\cdot a(M)\cdot (|M|\cdot(m+1))^{N-1}}\cdot b \\
	&\leq&(h+1)^{|T|\cdot(a(M)+1)\cdot(|M|\cdot (m+1))^{N-1}}\cdot b
\end{array}
}
\]
since $|T_i|\leq |T|$ and the nesting-depths of any of the $|T_i|$ is bounded by $a(T)-1$.
This completes the proof.
\end{proof}

Accordingly, the \emph{bit length} $Z(N)$ of numbers occurring in $\sem{t}$ for trees of depth $N$
is bounded by $O((|M|\cdot (m+1))^N)$ where $m$ is the maximal rank of an input symbol.
%
This means that, for $m>1$, the \emph{bit lengths} of occurring numbers can only be bounded by an exponential
in the sizes of $M$ and $A$. Still, in-equivalence can be decided in
\emph{randomized} polynomial time:

\begin{theorem}\label{t:nonnested-1}
In-equivalence of states of a non-self-nested total unary $y\DMTT$ transducer 
relative to a $\DTTA$ automaton,
is decidable in \emph{randomized polynomial time}, 
i.e.,
there is a polynomial probabilistic algorithm which in case of equivalence, always returns 
\textbf{false}, 
while in case of non-equivalence returns \textbf{true} with probability at least $0.5$.
\end{theorem}
\begin{proof}
Assume that $M$ is a non-self-nested total unary $y\DMTT$ transducer and 
$A$ a $\DTTA$ automaton. Our goal is to decide
whether or not $\sem{q_0}_M(t)=\sem{q'_0}_M(t)$ for all $t\in\L(A)$.
%
%
Now let $k$ denote any prime number. Then the set of integers modulo $k$, ${\mathbb Z}_k$, again forms a field.
This means that we can realize the algorithm for determining affine closures of the sets $V_p$ as well as
the check whether an affine mapping $H$ returns 0 for all elements of an affine basis now over ${\mathbb Z}_k$.
The resulting algorithm allows us to decide whether the outputs for $q_0,q_0'$ coincide for all
inputs from $\L(A)$ modulo the prime number $k$ by using polynomially many operations on numbers of length
$O(\log(k))$ only. In particular, if non-equivalence is found, then $q_0,q_0'$ cannot be equivalent relative
to $A$ over ${\mathbb Q}$ either.

Let ${2^D}$ be an upper bound to $Z(n\cdot(l+1)+1)\cdot h)$ ($n,l$ the number of states of $M$ and the number of 
parameters of states of $M$, respectively, and $h$ the number of states of $A$)
where $D$ is polynomial in the sizes of $M$ and
$A$. Then we have:

\begin{lemma}\label{l:aux}
$q_0,q_0'$ are equivalent relative to $A$ iff 
$q_0,q_0'$ are equivalent relative to $A$ modulo $2^D$ distinct primes.
\end{lemma}

\begin{proof}
Assume that the latter holds. Then the product already of the smallest $2^D$ primes vastly exceeds 
$2^{2^D}$. Therefore by the Chinese remainder theorem, $H(\sem{t}) = 0$ holds also over ${\mathbb Q}$
for all $t\in\L(A)$ of depth at most $(n\cdot(l+1)+1)\cdot h$. Therefore, $q_0,q'_0$ must be equivalent.
\end{proof}

\noindent
Clearly, if $q_0$ and $q_0'$ are equivalent relative to $\L(A)$, then they are also
equivalent relative to $\L(A)$ modulo every prime number $k$.
Therefore now assume that $q_0$ and $q_0'$ are not equivalent relative to $\L(A)$.
Let $K$ denote the set of all primes $k$ so that $q_0$ and $q_0'$ are still equivalent relative to $\L(A)$ modulo $k$.
By Lemma~\ref{l:aux}, this set has less than $2^D$ elements. 
Now consider the interval $[0,D\cdot e^{D}]$. 
Note that each number in this range has polynomial length only.
When $D$ is suitably large, this interval contains at least
$e^{D}\geq 4\cdot 2^D$ prime numbers (see, e.g., \cite{Hardy:Wright}). 
Therefore, with probability at least $0.75$, a prime randomly drawn from this range
is not contained in $K$ and therefore witnesses that $q_0$ and $q_0'$ are not equivalent 
relative to $\L(A)$.
Since a random prime can be drawn in polynomial time with probability $0.75$, and $0.75\cdot 0.75\geq 0.5$,
the assertion of the theorem follows.
%
\end{proof}

\section{General Unary Transducers with Parameters}\label{s:general}

\noindent
In the following, we drop the restriction that the $y\DMTT$ transducer 
$M$ is necessarily non-self-nested.
Then the polynomials ${\bf p}^{(f)}_{jk}$ are no longer necessarily multi-linear in the variables
${\bf x}_1,\ldots,{\bf x}_m$. 
Accordingly, 
techniques based on affine closures of the sets $\sem{{\sf dom}(p)}$ are no longer appropriate.

Instead, we propose to generally reason about \emph{properties} satisfied by the elements of the sets $\sem{\dom(p)}$.
Let ${\bf z} = \{{\bf z}_{qk}\mid q=1,\ldots,n, k=0,\ldots,l\}$ denote a fresh set of variables.
%
The key concept which we introduce here
is the notion of an \emph{inductive invariant} of $M$ relative to the $\DTTA$ automaton $A$.
As candidate invariants we only require conjunctions of equalities
$r \doteq 0$ where $r\in{\mathbb Q}[{\bf z}]$ is a polynomial over the variables ${\bf z}$ 
with rational coefficients. Instead of referring to such a conjunction directly, it is mathematically
more convenient to consider the \emph{ideal} generated from the polynomials in the conjunction.
Formally, an ideal of a ring $R$ is a subset $J\subseteq R$ such that for all $a,a'\in J,$ $a+a'\in J$
and for all $a\in J$ and $r\in R$, $r\cdot a\in J$. The smallest ideal containing a set $S$ of elements,
is the set $\angl{S} = \{\sum_{j=1}^k r_j\cdot s_j\mid k\geq 0, r_1,\ldots,r_k\in R, s_1,\ldots,s_k\in S\}$.
The smallest ideal containing ideals $J_1,J_2$ is their \emph{sum} $J_1+J_2= \{s_1+s_2\mid s_1\in J_1,s_2\in J_2\}$.

Using ideals instead of conjunctions of polynomial equalities is justified because for every 
$S\subseteq{\mathbb Q}[{\bf z}]$ and every
${\bf v}\in{\mathbb Q}^{n\times(l+1)}$, it holds that $s({\bf v})=0$ for all $s\in\angl{S}$ iff 
$s({\bf v})=0$ for all $s\in S$. 

An inductive invariant $I$ of the $y\DMTT$ transducer $M$ relative to $A$ 
is a family of ideals $I_{p}\subseteq {\mathbb Q}[{\bf z}],{p}\in P$,
such that for all transitions $\rho(p,f)=(p_1,\ldots,p_m)$,
\begin{eqnarray}
I_{p}	&\subseteq& 
	\{ r'\in{\mathbb Q}[{\bf z}]\mid r'[{\bf r}^{(f)}/{\bf z}]\in
	\angl{I_{p_1}({\bf x}_1)\cup\ldots\cup I_{p_m}({\bf x}_m)}\}
\label{eq:pol_inv}
\end{eqnarray}
holds where we used $[{\bf v}/{\bf z}]$ to denote the substitution
of the expressions ${\bf v}_{jk}$ for the variables ${\bf z}_{jk}$.
Likewise for an ideal $J\subseteq{\mathbb Q}[{\bf z}]$, 
$J({\bf x}_i)$ denotes the ideal:
\[
J({\bf x}_i) = \{s[{\bf x}_i/{\bf z}]\mid
	s\in J\}.
\]
The constraint \eqref{eq:pol_inv} for the transition $\rho(p,f) =(p_1,\ldots,p_m)$ 
formalizes the following intuition. For every polynomial $r'$, the polynomial
$r'[{\bf r}^{(f)}/{\bf z}]$ can be understood as the \emph{weakest precondition} of $r'$ w.r.t.\ the semantics
${\bf r}^{(f)}$ of the input symbol $f$. 
It is a polynomial in the variables ${\bf x}$ where
the variables in ${\bf x}_i$ refer to the $i$th argument of $f$.
The constraint \eqref{eq:pol_inv} therefore expresses that the weakest precondition of every polynomial in $I_{p}$
can be generated from the polynomials provided by $I$ for the states $p_i$ --- after the variables ${\bf z}$ 
therein have been appropriately renamed with ${\bf x}_i$.

We verify for every inductive invariant $I$ that each polynomial in the ideal $I_{p}$ constitutes a valid 
property of all input trees in ${\sf dom}({p})$. This means:

\begin{theorem}\label{t:inductive}\label{t:sound}
Assume that $I$ is an inductive invariant of the $y\DMTT$ transducer $M$
relative to $A$. Then for every state ${p}$ of $A$ and polynomial $r'\in I_{p}$,
$r'(\sem{t}) = 0$ holds for all $t\in{\sf dom}(p)$.
\end{theorem}
\begin{proof}
By structural induction on $t$, we prove that $r'(\sem{t})=0$ holds. 
Assume that 
$\rho(p,f) = (p_1,\ldots,p_m)$ and $t=f({\bf t}_1,\ldots,{\bf t}_m)$
where (by induction hypothesis) for every $i=1,\ldots,m$ and every $r'\in I_{p_i}$,
$r'(\sem{{\bf t}_i}) = 0$ holds.
Since $\sem{t}_{qk} = {\bf r}^{(f)}_{qk}(\sem{{\bf t}_1},\ldots,\sem{{\bf t}_m})$,
we have that 
\[
r'(\sem{t}) = r'[{\bf r}^{(f)}/{\bf z}] (\sem{{\bf t}_1},\ldots,\sem{{\bf t}_m})
\]
Since $I$ is inductive, 
$r'[{\bf r}^{(f)}/{\bf z}]$ can be rewritten as a sum:
\[
r'[{\bf r}^{(f)}/{\bf z}] = 
    \sum_{i=1}^m \sum_{\mu_i=1}^{u_i} r'_{i\mu_i} s_{i\mu_i}[{\bf x}_i/{\bf z}]
\]
for suitable polynomials $r'_{i\mu_i}$ where
for $i=1,\ldots,m$, all $s_{i\mu_i}\in I_{p_i}$.
Therefore for all $\mu_i$, $s_{i\mu_i}(\sem{{\bf t}_i}) = 0$, and thus 
\[
r'(\sem{t}) = r'[{\bf r}^{(f)}/{\bf z}] (\sem{{\bf t}_1},\ldots,\sem{{\bf t}_m}) = 0.
\]
\end{proof}

\noindent
We conclude that every inductive invariant $I$ with $r'\in I_{p}$
provides us with a \emph{certificate} that $r'(\sem{t}) = 0$ holds for all $t\in{\sf dom}(p)$.
In the next step, we convince ourselves that the reverse implication also holds, i.e.,
for all polynomials $r'$ for which $r'(\sem{t}) = 0$ holds for all $t\in{\sf dom}(p)$,
an inductive invariant $I$ exists with $r'\in I_{p}$. In order to prove this
statement, we consider the family $\bar I$ of ideals $\bar I_{p}, p\in P$, where
\[
\bar I_{p} = \{r'\in{\mathbb Q}[{\bf z}]\mid \forall\,t\in{\sf dom}(p).\;r'(\sem{t})=0\}.
\]
Thus, $\bar I_{p}$ is the set of \emph{all} polynomials which represent a polynomial property of trees
in ${\sf dom}(p)$.
We next prove that $\bar I$ is indeed an inductive invariant.

\begin{theorem}\label{t:complete}
${\bar I}$ is an inductive invariant of the $y\DMTT$ transducer $M$ relative to $A$. 
\end{theorem}

\begin{proof}
For any set $V\subseteq{\mathbb Q}^{n\times(l+1)}$ of vectors, 
let ${\cal I}(V)$ denote the set of polynomials $r'$ over ${\bf z}$
which vanish on $V$, i.e., with $r'({\bf v}) = 0$ for all ${\bf v}\in V$. We remark that
for disjoint sets
of variables ${\bf x}_1,\ldots,{\bf x}_m$ with ${\bf x}_i=\{{\bf x}_{ijk}\mid j=1,\ldots,n, k=0,\ldots,l\}$,
and arbitrary sets $V_i\subseteq {\mathbb Q}^{n\times(l+1)}$,
\begin{eqnarray}
{\cal I}(V_1\times\ldots\times V_m) =
\angl{{\cal I}(V_1)({\bf x}_1)\cup\ldots\cup{\cal I}(V_m)({\bf x}_m)}
\label{eq:carthesian}
\end{eqnarray}
holds when considered as ideals of ${\mathbb Q}[{\bf x}] = {\mathbb Q}[{\bf x}_1\cup\ldots\cup {\bf x}_m]$.
This means that the set of polynomials which vanish on the Cartesian product $V_1\times\ldots\times V_m$
is exactly given by the ideal in ${\mathbb Q}[{\bf x}]$ which is
generated by the polynomials in ${\mathbb Q}[{\bf x}_i]$ which vanish on the set $V_i$ ($i=1,\ldots,m$).
We remark that the ideal of ${\mathbb Q}[{\bf x}]$ generated from ${\cal I}(V_i)({\bf x}_i)$ is
exactly given by ${\cal I}(\top^{i-1}\times V_i\times \top^{m-i})$ where
$\top = {\mathbb Q}^{n\times(l+1)}$.
Accordingly, equality \eqref{eq:carthesian} can be rewritten to:
\[
{\cal I}(V_1\times\ldots\times V_m) = \sum_{i=1}^m {\cal I}(\top^{i-1}\times V_i\times \top^{m-i}).
\]
Thus, equality \eqref{eq:carthesian} is a consequence of the following lemma, which
we could not find in the literature. Although formulated for 
${\mathbb Q}$, the lemma holds (with the same proof) for any field.

\begin{lemma}\label{l:kemper}
Let $V_1\subseteq{\mathbb Q}^{m_1}, V_2\subseteq{\mathbb Q}^{m_2}$ be subsets of vectors, with $m_1,m_2$ positive integers.
Then 
\[
{\mathcal I}(V_1\times V_2) = {\mathcal I}(V_1\times{\mathbb Q}^{m_2}) + {\mathcal I}({\mathbb Q}^{m_1}\times V_2).
\]
\end{lemma}
%

\begin{proof} 
  Since $V_1 \times V_2 \subseteq V_1 \times \mathbb{Q}^{m_2}$, it
  follows that $I_1 := {\mathcal I}(V_1\times{\mathbb Q}^{m_2})
  \subseteq {\mathcal I}(V_1\times V_2)$. Likewise, $I_2 := {\mathcal
    I}({\mathbb Q}^{m_1}\times V_2) \subseteq {\mathcal I}(V_1\times
  V_2)$, and the inclusion ``$\supseteq$'' follows.

  The proof of the reverse inclusion uses Gr\"obner bases (for basic
  notions and concepts on Gr\"obner bases see the textbook by Becker
  and Weisspfenning~\cite{weiss}).
  Let ${\bf x}=\{{\bf x}_1\,\ldots,{\bf x}_{m_1+m_2}\}$ a suitable
  finite set of variables. Fix a monomial ordering on the polynomial
  ring $\mathbb{Q}[{\bf x}]$.  With respect to this monomial ordering,
  let $G_1$, $G_2$ be Gr\"obner bases of $I_1$ and $I_2$,
  respectively. Clearly $G_1 \cup G_2$ generates the sum $I_1 +
  I_2$. Since $I_1$ is generated by polynomials in
  $\mathbb{Q}[\mathbf{x}_1, \ldots,\mathbf{x}_{m_1}]$, we have $G_1
  \subset \mathbb{Q}[\mathbf{x}_1, \ldots,\mathbf{x}_{m_1}]$, and also
  $G_2 \subset \mathbb{Q}[\mathbf{x}_{m_1+1},
  \ldots,\mathbf{x}_{m_1+m_2}]$. It follows by Buchberger's criterion
  (see~\cite[Section~5.5]{weiss}) that $G_1 \cup G_2$ is a Gr\"obner
  basis of $I_1 + I_2$. This implies that each polynomial $f \in
  \mathbb{Q}[\mathbf{x}]$ has a unique normal form $g :=
  \operatorname{nf}(f)$, which (by definition) has no monomial that is
  divisible by the leading monomial of any polynomial in $G_1 \cup
  G_2$, and which satisfies $f - g \in I_1 + I_2$. Moreover, if $f \in
  I_1 + I_2$ then $g = 0$.

  For the proof of the reverse inclusion, take $f \in
  \mathbb{Q}[\mathbf{x}]$ that does {\em not} lie in $I_1 + I_2$. So
  $g := \operatorname{nf}(f) \ne 0$. We have to show $f \notin
  {\mathcal I}(V_1\times V_2)$, so we have to find ${\bf v} \in V_1$ and ${\bf w}
  \in V_2$ such that $f({\bf v},{\bf w}) \ne 0$. Considered as a polynomial in the
  variables $\mathbf{x}_1,\ldots,\mathbf{x}_{m_1}$, $g$ has a nonzero
  term $c \mathbf{x}_1^{e_1} \cdots \mathbf{x}_{m_1}^{e_{m_1}}$ with
  $c \in
  \mathbb{Q}[\mathbf{x}_{m_1+1},\ldots,\mathbf{x}_{m_1+m_2}]$. Since
  none of the monomials of~$c$ are divisible by any leading monomial
  of a polynomial from $G_2$, the Gr\"obner basis property of $G_2$
  implies $c \notin I_2$, so there exists ${\bf w} \in V_2$ such that $c({\bf w})
  \ne 0$.

  Now consider the polynomial $g_{\bf w} :=
  g(\mathbf{x}_1,\ldots,\mathbf{x}_{m_1},{\bf w}) \in
  \mathbb{Q}[\mathbf{x}_1,\ldots,\mathbf{x}_{m_1}]$. This is nonzero
  since one of its coefficients, $c({\bf w})$, is nonzero. Moreover, no
  monomial from $g_{\bf w}$ is divisible by any leading monomial of a
  polynomial from $G_1$, so $g_{\bf w} \notin I_1$, implying that there is a
  vector ${\bf v} \in V_1$ with $g_{\bf w}({\bf v}) \neq 0$.  This means $g({\bf v},{\bf w}) \ne
  0$. But $(f - g)({\bf v},{\bf w}) = 0$ since $f - g \in I_1 + I_2 \subseteq
  {\mathcal I}(V_1\times V_2)$, so we obtain $f({\bf v},{\bf w}) \ne 0$, finishing
  the proof. 
\end{proof}


\noindent
For each state ${p}$ of $A$, let $V_{p} =\{\sem{t}\mid t\in {\sf dom}({p})\}$.
Then the ideal $\bar I_{p}$ is exactly given by $\bar I_{p}={\cal I}(V_{p})$.
Assume that $r'\in\bar I_{p}$ and $\rho({p},f) = (p_1,\ldots,p_m)$ holds. Then for all tuples of trees 
$({\bf t}_1,\ldots,{\bf t}_m)$ with ${\bf t}_i\in{\sf dom}(p_i)$ ($i=1,\ldots,m$), 
$
r'(\sem{f({\bf t}_1,\ldots,{\bf t}_m)}) = 0
$
holds. Therefore,
\[
r'[{\bf r}^{(f)}/{\bf z}] ({\bf v}_1,\ldots,{\bf v}_m) = 0
\]
holds for all $({\bf v}_1,\ldots,{\bf v}_m)\in V_{p_1}\times\ldots\times V_{p_m}$.
Therefore,
\[
\begin{array}{lll}
r'[{\bf r}^{(f)}/{\bf z}]&\in&{\cal I}(V_{p_1}\times\ldots\times V_{p_m})	\\
	&=&
\angl{{\cal I}(V_{p_1})({\bf x}_1)\cup\ldots\cup{\cal I}(V_{p_m})({\bf x}_m)}\qquad
	\text{by \eqref{eq:carthesian}}	\\
	&=&
\angl{{\bar I}({\bf x}_1)\cup\ldots\cup{\bar I}({\bf x}_m)}.\qquad
\end{array}
\]
As a consequence, $\bar I$ satisfies the constraints \eqref{eq:pol_inv}
and therefore is an inductive invariant of $M$ relative to $A$.
\end{proof}
The inductive invariant $\bar I$ is the \emph{largest} invariant and, accordingly, the \emph{greatest}
fixpoint of the inclusions \eqref{eq:pol_inv}. Since the set of polynomial ideals in ${\mathbb Q}[{\bf x}]$
has unbounded \emph{decreasing} chains, it is unclear whether $\bar I$ 
can be effectively computed.

Let us first 
consider the case where the input alphabet of $M$ (and thus also of $A$) is
monadic. Then
the variables from ${\bf z}$ can be reused for the copy ${\bf x}_1$
of variables for the first (and only) argument of $f\in\Sigma^{(1)}$, implying
that every polynomial ${\bf r}^{(f)}_{qk}$ can be considered as a constant or a polynomial again over
the variables ${\bf z}$.
Thus,
the constraints in \eqref{eq:pol_inv} to be satisfied by an inductive invariant $I$, can be simplified to:
\begin{eqnarray}
I_{p}	&\subseteq& 
	\{r'\in{\mathbb Q}[{\bf z}]\mid
		r'({\bf r}^{(b)})=0\}\qquad\text{if}\;\rho(p,b)=()
\label{eq:monadic_0}	\\
I_{p}	&\subseteq& 
	\{r'\in{\mathbb Q}[{\bf z}]\mid
		r'[{\bf r}^{(f)}/{\bf z}] \in I_{p_1}\}\;\text{if}\;\rho(p,f)=p_1
\label{eq:monadic}
\end{eqnarray}
According to the second constraint, the demand for an equality $r'\doteq 0$ to hold at $p$ 
is transformed by the monadic input symbol $f$
into the demand for the equality $r'[{\bf r}^{(f)}/{\bf z}]\doteq 0$ to hold at $p_1$.
The propagation of these demands generated for the equality $H \doteq 0$ to hold at $p_0$
can be expressed by the following system of constraints:
\begin{eqnarray}
I_{p_0} &\supseteq&\angl{H} \nonumber		\\
I_{p_1}	&\supseteq&\{r'[{\bf r}^{(f)}/{\bf z}]\mid r'\in  I_{p}\}
\qquad\text{if}\; \rho(p,f)=p_1
\label{eq:wp}
\end{eqnarray}
Recall that Hilbert's basis theorem implies that each ideal $J\subseteq{\mathbb Q}[{\bf z}]$
can be represented by a finite set of polynomials $s_1,\ldots,s_u$ so that 
$J=\angl{s_1,\ldots,s_u}$ and likewise, that each \emph{increasing} chain $J_0\subseteq J_1\subseteq \ldots$ of ideals
is ultimately stable. Therefore, the system \eqref{eq:wp} has a \emph{least solution}, 
which is attained after finitely many fixpoint iterations.
We claim:

\begin{lemma}\label{l:wp-correctness}
Assume that $I$ is the least solution of the system of constraints \eqref{eq:wp}.
Then $I$ is an inductive invariant iff for each transition
$\rho(p,b) = ()$ of $A$, $r'({\bf r}^{(b)})=0$ for all $r'\in I_{p}$.
In this case, it is the \emph{least} inductive invariant $I'$ with $H\in I'_{p_0}$.
Otherwise, no inductive invariant with this property exists.
\end{lemma}

\begin{proof}
We have that $I$ is a solution of \eqref{eq:wp} iff $I$ satisfies
the constraints 
\eqref{eq:monadic}.
Moreover,
$r'({\bf r}^{(b)})=0$ for all $r'\in I_{p'}$ holds for all 
$\rho({p},b) = ()$ of $A$ iff 
$I$ satisfies the constraints \eqref{eq:monadic_0}. 
Therefore, $I$ is an inductive invariant with $H\in I_{p_0}$,
iff these two assumptions are met.

Now assume that $I$ is the least solution of \eqref{eq:wp}.
If it passes all tests on the transitions $\rho(p,b)$,
it therefore must be the least inductive invariant $I'$ with
$H\in I'_{p_0}$.
If it does not pass all tests, then there cannot be any inductive
invariant $I'$ with $H\in I'_{p_0}$. This can be seen as follows. 
Assume for a contradiction that there is an inductive invariant $I'$ 
with $H\in I'_{p_0}$. Since $I'$ satisfies the constraints in \eqref{eq:monadic},
$I'$ is also a solution of \eqref{eq:wp}. Therefore, $I_{p}\subseteq I'_{p}$
for all states ${p}$ of $A$. Now since already $I$ does not pass all tests
on transitions $\rho({p},b)=()$, then $I'$ cannot pass all these tests either.
But then $I'$ does not satisfy the constraints 
\eqref{eq:monadic} and therefore fails to be an inductive invariant --- contradiction.
\end{proof}

\noindent
Since the least solution of system \eqref{eq:wp} can effectively be computed and 
the tests required by Lemma~\ref{l:wp-correctness} can also be effectively performed,
we obtain:

\begin{theorem}
For a total unary $y\DMTT$ transducer $M$ with a monadic input alphabet, it is decidable
whether or not two states 
are equivalent relative to a $\DTTA$ automaton $A$.
\end{theorem}

\begin{proof}
Let $H$ denote the polynomial ${\bf z}_{j_00}-{\bf z}_{j'_00}$.
Then $H(\sem{t'})=0$ for all $t'\in{\sf dom}(p_0)$ holds iff $H\in \bar I_{p_0}$.
Now, $H\in I_{p_0}$ for some inductive invariant $I$ iff 
$H\in I'_{p_0}$ for the least inductive invariant $I'$ which,
by Lemma~\ref{l:wp-correctness} can be effectively computed. 
Since membership of a polynomial in an ideal can be effectively decided,
the claim of the theorem follows.
\end{proof}


\noindent
In the following, we finally drop also the assumption that
the $y\DMTT$ transducer 
has a monadic input alphabet.
What we keep is the assumption that the output alphabet is unary.
For this case, we prove that equivalence is still decidable. 
An indicator for the extra complication due to non-monadic input symbols
is that we are only able to provide two \emph{semi}-algorithms,
one which provides a proof of equivalence if equivalence holds, and another which provides an input tree
for which the output differ --- whenever non-equivalence holds.


\noindent
In case of non-monadic input symbols, it is no longer clear whether
computing the greatest solution of constraint system \eqref{eq:pol_inv} can be replaced by
computing the least solution over some suitably defined alternative constraint system over ideals.
What we still know is that every ideal of ${\mathbb Q}[{\bf z}]$ can be represented 
by a finite set of polynomials with coefficients, which can be chosen from ${\mathbb Z}$. 
Since the validity of the inclusions \eqref{eq:pol_inv} can be
effectively decided for any given candidate invariant $I$, 
the set of \emph{all} inductive invariants of $M$ relative to $A$
is recursively enumerable. Accordingly,
if ${\bf z}_{j_00}-{\bf z}_{j'_00} = 0$ holds for all vectors $\sem{t}, t\in{\sf dom}(p_0)$, 
an inductive invariant certifying this fact, will eventually be found in the enumeration.
In this way, we obtain a \emph{semi}-decision procedure for equivalence of the states $j_0,j'_0$ of
$M$ relative to $A$.
The fact, on the other hand, that ${\bf z}_{j_00}-{\bf z}_{j'_00} = 0$ does not hold for all 
$\sem{t}, t\in{\sf dom}(p_0)$, 
is witnessed by a specific tree $t\in{\sf dom}(p_0)$ for which
$\sem{t}_{j_00} -\sem{t}_{j'_00}\neq 0$.
Since ${\sf dom}(p_0)$ is recursively enumerable as well, 
we obtain another \emph{semi}-decision procedure, now for non-equivalence of states $j_0,j'_0$ of
$M$ relative to $A$. 
Putting these two \emph{semi}-decision procedures together, we obtain:

\begin{theorem}\label{t:general}
Assume that $M$ is a total $y\DMTT$ transducer with unary output alphabet, $A$ is a $\DTTA$ automaton.
Then it is decidable whether or not two states $j_0,j'_0$ of $M$ 
are equivalent relative to $A$.
\end{theorem}

\noindent
In the same way as in the last section, Theorem~\ref{t:general} provides us with a decision procedure
for possibly partial unary $y\DMTT$ transducers. We obtain our main technical result:

\begin{theorem}\label{t:partial-1}
Equivalence for (possibly partial) unary $y\DMTT$ transducers is decidable.
\end{theorem}

\section{A More Practical Algorithm}\label{s:practical}\label{s:algo}

Clearly, checking all input trees is perhaps
not the most systematic way of identifying a counter example to equivalence.
Likewise, enumerating all mappings $p\mapsto{\cal I}_p$, in quest for a sufficiently strong
inductive invariant seems completely impossible to be turned into a practical algorithm.
Therefore, in this section we provide more realistic implementations of the two semi-algorithms
to decide equivalence.

To accomplish the task of identifying counter examples, we take a closer look at the
greatest fixpoint iteration to determine the greatest inductive invariant $p\mapsto\bar{\cal I}_p$.
For $p\in P$, let $\bar{\cal I}_p^{(0)} =\angl{1}_{\Q[\bz]} = \Q[z]$, i.e., the full polynomial 
ring, and for $d>0$, define $\bar{\cal I}_p^{(d)}$ as
\begin{eqnarray}
\bar{\cal I}_p^{(d)} &=& \bigcap\{\sem{f}^\sharp(\bar{\cal I}_{p_1}^{(d-1)},\ldots,
						 \bar{\cal I}_{p_k}^{(d-1)})\mid
				\rho(p,f)=p_1\ldots p_k\}\label{eq:iterate}	\\
\sem{f}^\sharp(I_1,\ldots,I_k)
	&=&\{ r\in\Q[\bz]_\mid r[r^{(f)}/\bz]\in
			\angl{I_1(\bx_1)\cup\ldots\cup I_k(\bx_k)}_{\Q[\bz]}\} \nonumber\\
	&=&(\angl{\bz_{qi}-r^{(f)}_{qi}\mid q=1,\ldots,n,i=0,\ldots,l}_{\Q[\bz]}\;\oplus\nonumber\\
	& &\qquad 	\angl{I_1(\bx_1)\cup\ldots\cup I_k(\bx_k)}_{\Q[\bz]})\cap\Q[\bz]
				\label{eq:explicit}
\end{eqnarray}
For every $p\in P$ and $d\geq 0$, let ${\sf dom}_d(p)$ denote the set of all input trees 
$t\in{\sf dom}(p)$ of depth at most $d$ where we consider leaves to have depth 1.
Then we have:

\begin{lemma}\label{l:counterexample}
\begin{enumerate}
\item	For every $d\geq 0$,
	$\bar{\cal I}_p^{(d)} = \{r\in\Q[\bz]\mid
			\forall\,t\in{\sf dom}_k(p).\;r(\sem{t}) = 0\}$;
\item	For every $p\in P$, $\bar{\cal I}_p = \bigcap\{\bar{\cal I}_p^{(d)}\mid d\geq 0\}$.
\end{enumerate}
\end{lemma}

\begin{proof}
Statement 1 follows by induction on $d$ along the same lines as the proofs of 
Theorems~\ref{t:sound}
and~\ref{t:complete}.
Statement 2 follows from statement 1 as the intersection to the right consists of all
polynomials $r\in\Q[\bz]$ so that $r(\sem{t}) = 0$ for all $t\in{\sf dom}(p)$ ---
which precisely is the definition of $\bar{\cal I}_p$.
\end{proof}

\noindent
From statement 1 of Lemma~\ref{l:counterexample}, we conclude that
there is a counter example to equivalence of $q,q'$ of depth $d$ iff 
$H_{q,q'}\not\in\bar{\cal I}_{p_0}^{(d)}$ for the initial state $p_0$ of the $\DTTA$ automaton $B$.
Thus, the semi-algorithm for falsifying equivalence, can be formulated as:
\[
\begin{array}{l}
{\bf for}\;(d\geq 0)\;\{	\\
\qquad \text{determine}\,p\mapsto\bar{\cal I}_p^{(d)};\\
\qquad{\bf if}\;(H_{q,q'}\not\in\bar{\cal I}_{p_0}^{(d)})\;
		{\bf return}\;\text{``not equivalent''};\\
\qquad\}
\end{array}
\]
We turn to the efficient enumeration of candidate invariants.
Let us again fix some bound $d$. This time, the bound $d$ is used as the degree bound
to the polynomials to be considered during the fixpoint iteration. 
Let $\Q_d[\bz]$ denote the set of all polynomials in $\Q[\bz]$ of total degree at most $d$.
Here, the total degree of a polynomial $r$ is the maximal sum of exponents of a monomial
occurring in $r$.
For an ideal ${\cal I}\subseteq\Q[\bz]$, let us denote ${\cal I}_d$ as the intersection
${\cal I}\cap\Q[\bz]$. This set of polynomials can be considered as a vector space of finite 
dimension and can also be considered as a \emph{pseudo ideal} in the sense of Colon \cite{Colon07}.

The intersection ${\cal I}_d$ can be effectively computed as follows. 
Given a Gr\"obner basis $G$ for ${\cal I}$ relative to some graded lexicographical ordering
of monomials,
it suffices to extract the subset $G_d$ of polynomials in $G$ of total degrees bounded by
$d$. Then a set of generators of ${\cal I}_d$ considered as a vector space is given by 
the set of polynomials 
$g\cdot m$ with $g\in G_d$ for monomials $m$ 
with ${\sf def}(m)+{\sf deg}(g)\leq d$.
Moreover, the ideal $\angl{G_d}_{\Q[\bz]}$ is the smallest ideal ${\cal I}'\subseteq{\cal I}$
so that ${\cal I}'\cap\Q_d[\bz] = {\cal I}\cap\Q[\bz]$.

Let $\alpha_d$ denote the function that maps each ideal ${\cal I}$ to the corresponding ideal 
$\angl{G_d}_{\Q[\bz]}$.
Let ${\mathbb D}_d$ denote the set of all ideals generated from Gr\"obner bases of total degree
at most $d$. In this complete lattice, all decreasing sequences are ultimately stable.
The idea is to compute increasingly precise abstractions of
the mapping $p\mapsto\bar{\cal I}_p$ by means of the complete lattices ${\mathbb D}_d$.
For $d\geq 1$, we put up the constraint system over ${\mathbb D}_d$, consisting of all
constraints
\begin{eqnarray}
I_p	&\subseteq &
	\alpha_d(\sem{f}^\sharp(I_{p_1},\ldots,I_{p_k})
			\label{eq:lower}
\end{eqnarray}
for every transition $\rho(p,f)=p_1\ldots p_k$.
Since all right-hand sides are monotonic, the greatest solution of this system exists.
Since ${\mathbb D}_d$ has finite descending chains only, the greatest solution is
attained after finitely many fixpoint iterations. 
Let $p\mapsto{\cal I}_{p,d}$ denote the resulting greatest solution.
We have:
	
\begin{lemma}\label{l:lower}
\begin{enumerate}
\item	For all $d\geq 1$, $p\mapsto{\cal I}_{p,d}$ is an inductive invariant;
\item	For all $d\geq 1$, 
	${\cal I}_{p,d}\subseteq {\cal I}_{p,d+1}$ holds for every $p\in P$;
\item	There exists some $d\geq 1$ such that
	${\cal I}_{p,d} = \bar{\cal I}_p$ for all $p\in P$.
\end{enumerate}
\end{lemma}

\begin{proof}
The first two statements are obvious. 
%
%
In order to prove the third statement, consider the greatest inductive
invariant $p\mapsto\bar{\cal I}_p$. Thus in particular, ${\cal I}_{p,d}\subseteq\bar{\cal I}_p$
for all $p$ and $d$. For every state $p$, let $G_p$
denote the Gr\"obner base of $\bar{\cal I}_p$. Let $\bar d$ denote the maximal
global degree of any polynomial in the set $\bigcup\{G_p\mid p\in P\}$.
Then $p\mapsto\bar{\cal I}_p$ is a solution of the constraint system \eqref{eq:lower}.
Accordingly, $\bar{\cal I}_p\subseteq{\cal I}_{p,\bar d}$. Hence $\bar{\cal I}_p={\cal I}_{p,\bar d}$, and statement 3 follows.
\end{proof}

\noindent
In light of the argument for proving statement 3 of Lemma~\ref{l:lower}, we observe that for any $d$, 
$p\mapsto{\cal I}_{p,d}$ represents the largest inductive invariant which can be represented by
a Gr\"obner basis with maximal total degree $d$.
Given Lemma~\ref{l:lower}, the semi-algorithm for verifying equivalence therefore 
looks as follows:
\[
\begin{array}{l}
{\bf for}\;(d\geq 1)\;\{        \\
\qquad \text{determine}\,p\mapsto{\cal I}_{p,d};\\
\qquad{\bf if}\;(H_{q,q'}\in{\cal I}_{p_0,d})\;
                {\bf return}\;\text{``equivalent''};\\
\qquad\}
\end{array}
\]
This algorithm now provides a systematic way to generate 
inductive invariants of increasing precision thus complementing 
the systematic enumeration method of counter examples
of increasing depths.

\section{From yDT transducers to unary yDMTT transducers}\label{s:sim}\label{s:mtt}

\noindent
In the following, we show that every total $y\DT$ transducer can be simulated by
a total $y\DMTT$ transducer with a unary output alphabet and polynomial size.
This is the content of the next lemma. Assume that the 
output alphabet 
is given by $\Delta = [s]$. 
By considering the elements of $\Delta$ as non-zero digits of the number
system with base $s+1$, each element $w\in\Delta^*$ can be uniquely represented by
a natural number.
If $w=w_1\ldots w_k$, $w_j\in[s]$, this number is given by
$[w]_{s+1} = \sum_{j=1}^k w_j\cdot (s+1)^j$.
In particular, $[\epsilon]_{s+1} = 0$, i.e., the empty string is represented by 0.
We have:

\begin{lemma}\label{l:sim}
Assume that $M$ is a total $y\DT$ transducer with set $[n]$ of states and output alphabet $[s]$. 
Then a unary $y\DMTT$ transducer $N$ with the same set of states and a single parameter,
can be constructed in polynomial time so that for every state $q\in[n]$ and input tree $t$,
$\sem{q}_N(t)(\epsilon)=[\sem{q}_M(t)]_{s+1}$.

Moreover, if $M$ is linear, then $N$ is non-self-nested.
\end{lemma}

\begin{proof}
Let $M=(Q,\Sigma,\Delta,q_0,R)$ where  $Q=[n]$.
We define $N=(Q,\Sigma,\{d\},q_0,R')$ as follows. For every
rule $q(f(x_1,\ldots,x_k))\to T$ in $R$ we let the rule 
$
q(f(x_1,\ldots,x_k),y_1)\to {\cal U}[T]
$
be in $R'$. The parameter $y_1$ 
is meant to contain 
the right context (in unary).
The mapping ${\cal U}[T]$ is defined as follows:
\[
\begin{array}{lll}
{\cal U}[aT]	&=& a + (s+1)\cdot{\cal U}[T]	\\
{\cal U}[q'(x_i)T] &=& q'(x_i,{\cal U}[T])	\\
{\cal U}[\epsilon]	&=&	y_1.
\end{array}
\]
%
%
For the $y\DMTT$ transducer $N$ we prove the following invariant:
\[
\begin{array}{lll}
\sem{q}_N(t)([w]_{s+1})&=& 
\relax[\sem{q}_M(t)\;w]_{s+1}.
\end{array}
\]
From that, the statement follows by choosing $q=q_0$ and $w=\epsilon$.
In order to prove the invariant, we proceed by induction on the structure of $t$.
So assume that $t=f({\bf t}_1,\ldots,{\bf t}_m),m\geq 0,$ where $\delta(q,f)=T$. 
By induction, we may assume that the invariant already holds for ${\bf t}_1,\ldots,{\bf t}_k$
and all output words $w$. Then we prove that for all subsequences $T'$ of $T$ and all words $w\in[s]^*$
the following invariant holds:
\[
\begin{array}{lll}
\sem{{\cal U}[T']}_N\; {\bf t}\;([w]_{s+1}) 
&=&	
\relax[\sem{T'}_M\; {\bf t}\;w]_{s+1}	
\end{array}
\]
where ${\bf t}=({\bf t}_1,\ldots,{\bf t}_m)$.
The invariant for $t$ follows because 
$\sem{q}_N\;(t)\;(y_1) = \sem{{\cal U}[T]}_N\;{\bf t}\;(y_1)$ and 
$\sem{q}_M(t) = \sem{T}_M\;{\bf t}$.
If $T'=\epsilon$, we have that
\[
\begin{array}{lllll}
\relax[\sem{\epsilon}_M\;{\bf t}\;w]_{s+1}  &=& \relax[w]_{s+1}	
				      &=& \sem{{\cal U}[\epsilon]}_N\;{\bf t}\;([w]_{s+1})	
\end{array}
\]
and the invariant holds.

If $T' = aT''$ for some output symbol $a\in[s]$, we have that
\[
\begin{array}{lll}
\relax[\sem{aT''}_M\;{\bf t}\;w]_{s+1}  &=& \relax[a\sem{T''}_M\;{\bf t}\;w]_{s+1}       \\
                                      &=& a+(s+1)\cdot[\sem{T''}_M\;{\bf t}\;w]_{s+1} 	\\
                                      &=& a+(s+1)\cdot\sem{{\cal U}[T'']}_N\;{\bf t}\;([w]_{s+1}) 	
				      
			\quad\text{by induction for}\; T''\\
                                      &=& \sem{{\cal U}[aT'']}_N\;{\bf t}\;([w]_{s+1}) 
\end{array}
\]
and the invariant holds, by induction, for $T''$.

Therefore, it remains to consider the case where $T'=q'(x_i)T''$. Then:
\[
\begin{array}{lll}
\relax[\sem{q'(x_i)T''}_M\;{\bf t}\;w]_{s+1} &
=& \relax[\sem{q'}_M({\bf t}_i) \sem{T''}_M\;{\bf t}\;w]_{s+1}       \\
                                     & =& \sem{q'}_N({\bf t}_i)([\sem{T''}_M\;{\bf t}\;w]_{s+1})	
			\;\text{by induction for}\;{\bf t}_i\;\text{and}\; w'= \sem{T''}_M\;{\bf t}\;w	 \\
                                     &=& \sem{q'}_N({\bf t}_i)(\sem{{\cal U}[T'']}_N\;{\bf t}\;([w]_{s+1})) 
			\;\text{by induction for}\;T''	\\
                                      &=& \sem{{\cal U}[q'(x_i)T'']}_N\;{\bf t}\;([w]_{s+1}) 
\end{array}
\]
and the assertion follows.
--- Obviously, if $M$ is linear then $N$ is non-self-nested.
\end{proof}

\section{Output in the Free Group}\label{s:free}

In some applications, the output produced by a transducer cannot be considered
as a sequence of uninterpreted letters, but consists in
a sequence of operators. A first step in direction of dealing with such
interpreted output is to consider the output to be an element of a 
\emph{free group}. In that, we go beyond the free monoid as output domain
and assume
that for each output letter $a$, there is an \emph{inverse}
letter $a^-$ together with the cancellation rules 
\[
a a^- = a^- a = \epsilon\,.
\]
And we ask whether equivalence of transducers remains decidable if output strings
are considered equivalent up to applications of these rewrite rules.

Let us first consider output generated in the free group ${\cal F}_1$ with a single generator.
%
The free group ${\cal F}_1$ is isomorphic to the integral ring ${\mathbb Z}$.
This means that our construction for $y\DMTT$ transducers with unary output alphabet can be readily 
applied also to $y\DMTT$ transducers with output in ${\cal F}_1$.
We obtain:

\begin{theorem}\label{t:free-1}
Assume that $M$ is a total $y\DMTT$ transducer 
with output in the free group ${\cal F}_1$, and $A$ is a $\DTTA$ automaton.
Then it is decidable whether or not two states $j_0,j'_0$ of $M$
are equivalent relative to $A$.
If $M$ is non-self-nested, in-equivalence can even be decided in randomized polynomial time.
\qed
\end{theorem}

\noindent
We now consider the case when we have two distinct output symbols $a,b$ and together with their respective
inverses $a^-$ and $b^-$, i.e., we consider outputs in the free group ${\cal F}_2$
with two generators.
We do not know how the result for ${\cal F}_1$ and $y\DMTT$ transducers can be generalized to this 
more general situation.
What we can do, however, is to consider $y\DT$ transducers where output symbols 
are interpreted as $l\times l$ matrices with entries in $\Q$.
Let ${\cal M}_l(\Q)$ denote the monoid of all such matrices where the monoid operation
is matrix multiplication. Given an interpretation $\alpha:\Delta\to{\cal M}_l(\Q)$ of output symbols,
every output string $w$ then represents a matrix $\alpha(w)\in{\cal M}_l(\Q)$.
Accordingly, the semantics $\sem{t}$ of an input tree $t$ turns into a vector of matrices from ${\cal M}_l(\Q)$, i.e., 
$\sem{t}\in \Q^{n\cdot(l\times l)}$. Moreover, every input symbol $f$
of arity $k\geq 0$ corresponds to a mapping:
\[
\sem{f}:\Q^{n\cdot(l\times l)}\times\ldots\times\Q^{n\cdot(l\times l)}\to\Q^{n\cdot(l\times l)}
\]
($n$ the number of states of the transducer)
which transforms the vectors corresponding to the semantics of the argument trees into
the vector corresponding to the result vector of matrices for the whole input tree.
By induction on the structure of right-hand sides, we find that the entry
$(\sem{f}(\bx_1,\ldots,\bx_k))_{q\lambda\mu}\,{=:}\,r^{(f)}_{q\lambda\mu}$ 
is a polynomial in the variables $\bx_{iq'\lambda'\mu'}$,
$i\in\{1,\ldots,k\}, q'\in\{1,\ldots,n\},\lambda',\mu'\in\{1,\ldots,l\}$.
In the particular case that the $y\DT$ transducer 
is linear, each $r^{(f)}_{q\lambda\mu}$ is multi-affine
in the vectors $\bx_1,\ldots,\bx_k$.
\begin{example}\label{e:matrix}
Consider the two matrices
\[
m_1=\left[\begin{array}{ll}
	3&1\\
	0&1
	\end{array}\right]\qquad\qquad
m_2=\left[\begin{array}{ll}
	3&2\\
	0&1
	\end{array}\right]
\]
Let $\Delta=\{a_1,a_2\}$.
The monoid homomorphism $\alpha:\Delta^*\to{\cal M}_2(\Q)$ mapping $a_i$ to $m_i$ is injective and
given by 
\[
\alpha(a_{j_1}\ldots a_{j_s}) =
	\left[\begin{array}{ll}
	3^s& w\\
	0&1
	\end{array}\right]\qquad\text{where}\qquad w = \sum_{\lambda=0}^{s-1} 3^\lambda\cdot j_\lambda
\]
In this way, the free monoid of strings over the alphabet $\Delta$ can be
represented by the sub-monoid of ${\cal M}_2(\Q)$ generated by $m_1,m_2$.
Consider, e.g., a transition
\[
q(f(x_1,x_2)) \longrightarrow a_1\, q'(x_2)\, a_2\, q''(x_1) 
\]
of some deterministic  $y\DT$ transducer with output in ${\cal M}_2(\Q)$ according to the given $\alpha$.
Then according to our definition,
\[
\begin{array}{lll}
(\sem{f}(\bx_1,\bx_2))_q	&=& m_1\cdot\bx_{2q'}\cdot m_2\cdot \bx_{1q''}	\\[1ex]
&=&
	\left[\begin{array}{ll}
	3&1\\
	0&1
	\end{array}\right] \cdot
	\left[\begin{array}{ll}
	\bx_{2q'11}&\bx_{2q'12}\\
	\bx_{2q'21}&\bx_{2q'22}
	\end{array}\right] \cdot
	\left[\begin{array}{ll}
	3&2\\
	0&1
	\end{array}\right] \cdot
	\left[\begin{array}{ll}
	\bx_{1q''11}&\bx_{1q''12}\\
	\bx_{1q''21}&\bx_{1q''22}
	\end{array}\right] 
\end{array}
\]
\qed
\end{example}

Let ${\cal H}_{j_0j'_0}$ denote the set of polynomials $\bz_{j_0\lambda\mu}-\bz_{j_0'\lambda\mu},
\lambda,\mu\in\{1,\ldots,l\}$. Then the states $j_0,j_0'$ are equivalent relative to the
$\DTTA$ automaton $A$ iff $r(\sem{t}) = 0$ for all $r\in{\cal H}_{j_0j'_0}$ and $t\in{\sf dom}(p_0)$.
In general, this can be decided by the algorithm which we applied in the proof of 
Theorem~\ref{t:general},
but adapted to the new construction of the polynomials $r^{(f)}_{q\lambda\mu}$ and
where the single linear target equality $H_{j_0j'_0}$ is replaced with the 
conjunction of the finitely many target equalities from ${\cal H}_{j_0j'_0}$.
In the case of linear $y\DT$ transducers, the algorithm from Section~\ref{s:nonnested} can be adapted 
accordingly. Therefore, we obtain:

\begin{theorem}\label{t:matrix}
Assume that $M$ is a total $y\DT$ transducer with output in the monoid ${\cal M}_l(\Q)$, 
and $A$ is a $\DTTA$ automaton.
Then it is decidable whether or not two states $j_0,j'_0$ of $M$
are equivalent relative to $A$.
If the $y\DT$ transducer is linear and the ranks of the input symbols is bounded,
in-equivalence is decidable in randomized polynomial time.
\qed
\end{theorem}

\noindent
We remark that extending the matrix monoid 
considered in Example~\ref{e:matrix} with inverses,
would not provide us with a free group.
Instead, however, we can 
consider the subgroup of $2\times 2$ matrices generated from the elements:
\[
a=\left[\begin{array}{ll}
	1&0\\
	2&1
	\end{array}\right]\qquad\qquad
b=\left[\begin{array}{ll}
	1&2\\
	0&1
	\end{array}\right]
\]
This group is also known as \emph{Sanov} group $\cal S$ (see, e.g., Example 4.5.1 of 
\cite{lecture-notes}).
Since both matrices have determinant 1, all elements in $\cal S$ have integer coefficients
only. In particular, the inverses of the two generators are given by:
\[
a^-=\left[\begin{array}{rl}
	1&0\\
	-2&1
	\end{array}\right]\qquad\qquad
b^-=\left[\begin{array}{lr}
	1&-2\\
	0&1
	\end{array}\right]
\]
The Sanov group is particularly useful for us, since 
${\cal S}$ with subgroup generators $a$ and $b$ 
is isomorphic to the free group ${\cal F}_2$ 
freely generated from $a,b$.
Thus, our results on y$\DT$ transducers with output in matrix monoids in Theorem~\ref{t:matrix}
gives us:

\begin{theorem}\label{t:free-2}
Assume that $M$ is a total $y\DT$ transducer 
with output in the free group ${\cal F}_2$, and $A$ is a $\DTTA$ automaton.
Then it is decidable whether or not two states $j_0,j'_0$ of $M$
are equivalent relative to $A$.
If the $y\DT$ transducers are linear and the ranks of their input symbols are bounded,
in-equivalence is decidable in randomized polynomial time.
\qed
\end{theorem}

\noindent
Since the free group with two generators has a free group with $l\geq 2$ generators as a
subgroup, Theorem~\ref{t:free-2} implies that equivalence of total $y\DT$ transducers 
relative to some $\DTTA$ automaton
is also decidable when the outputs are in a free group with $l\geq 2$ generators.


\noindent
Lemma~\ref{l:sim} allows to apply our decision procedures for unary $y\DMTT$ transducers to decide
equivalence for $y\DT$ transducers with arbitrary output alphabets. 
%
%
Via Lemma~\ref{l:sim}, equivalence for linear (possibly partial) $y\DT$ transducers 
is reduced to equivalence of 
non-self-nested unary $y\DMTT$ transducers,
while equivalence for arbitrary (possibly partial and non-linear) $y\DT$ transducers is reduced to 
equivalence of general unary $y\DMTT$ transducers. In summary, we obtain:

%
%

\begin{theorem}\label{t:main}
Equivalence of arbitrary $y\DT$ transducers is decidable.
If the $y\DT$ transducers are linear and the ranks of their input symbols are bounded,
in-equivalence is decidable in randomized polynomial time.
\end{theorem}


\medskip

\noindent
The second part of Theorem~\ref{t:main} follows 
from Theorem~\ref{t:nonnested-1} and Lemma~\ref{l:bin}.
One particular subcase of Theorem~\ref{t:main} 
is when the input alphabet is monadic.
This case is known to be equivalent to 
the sequence equivalence problem of $\HDToL$ systems~
\cite{DBLP:journals/corr/Maneth14,DBLP:journals/tcs/Honkala00a}.
By Lemma~\ref{l:sim}, this problem 
can be reduced to the equivalence problem for unary $y\DMTT$ transducers with monadic output alphabet,
for which a \emph{direct} algorithm based on fixpoint iteration over polynomial ideals has been
presented in the last section.
The equivalence problem for $y\DT$ transducers with non-monadic input alphabets, as shown to be decidable
in Theorem~\ref{t:main}, seems to be significantly more difficult. 

\section{Application to other Types of Transducers}\label{s:results}

Tree transducers can be equipped with regular look-ahead.
For top-down transducers this increases the expressive power.
A top-down or macro tree-to-string transducer \emph{with regular look-ahead}
($y\DT^R$ and $y\DMTT^R$ transducer)
consists of an ordinary such transducer together with a complete
deterministic bottom-up tree automaton $B$. 
For a $y\DT^R$ transducer, 
a rule is of the form
\[
q(f(x_1:p_1,\dots,x_m:p_m))\to T
\] 
where $T$ is as for ordinary $y\DT$ transducers, and 
the $p_i$ are states of $B$. The rule is applicable to
an input tree $f\,\bt$ if $B$ arrives in state $p_i$ on input
tree $\bt_i$ for every $i\in\{1,\dots,m\}$.
Our result extends to look-ahead, using the technique as in~\cite{EngelfrietMS09}:
one changes the input ranked alphabet to contain state information of $B$,
and changes the transducer to check the correctness of the information.

By a result of~\cite{DBLP:journals/jcss/EngelfrietM02} the class of translations realized by 
$y\DT^R$ transducers is equal to 
the class realized by 
macro transducers which use each parameter in a rule precisely once (and have look-ahead).
In fact, by the results of~\cite{DBLP:journals/iandc/EngelfrietM99} we can state the result 
in terms of $y\DMTT^R$ transducers 
that are \emph{finite-copying in the parameters} ($y\DMTT_{\text{fcp}}^R$ transducers).
A $y\DMTT^R$ transducer is finite-copying in the parameters if  
there exists a $k$ such that for every input tree $s$, state $q$ (of rank $l+1$), 
and $j\in[l]$, the number of occurrences of $y_j$ in $\sem{q}(s)$ is $\leq k$.

\begin{corollary}\label{cor1}\label{c1}
Equivalence of $y\DT^R$ transducers and $y\DMTT^R_{\text{\rm fcp}}$ transducers is decidable.
\end{corollary}

A variation of transducers that has been considered in the context of \XML,
are transducers of \emph{unranked trees}.
In an unranked tree, the number of children of a node is not determined
by the label of that node, but is independent.
For instance the term $a(a(a,b),a,a,a)$ represents an unranked tree.
\XML document trees are naturally modeled by unranked trees. 

There are several models of top-down tree transducers for
unranked trees. 
In~\cite{persei04} macro forest transducers,
and their parameterless version, the \emph{forest transducers}, are defined.

The rules of a forest transducer are very similar to the rules of 
a $y\DT$ transducer and are of the form
$q(a(x_1,x_2)\to T$ where $T$ is a string as in a $y\DT$ transducer,
with the only difference that $T$ may contain special symbols
``('' and ``)'' of opening and closing brackets, and if so, then
$T$ must be well-balanced with respect to these brackets.
If such a rule is applied to an unranked $a$-labeled node $u$,
then $x_1$ represents the first subtree of $u$, and $x_2$ represents
the next sibling of $u$. The special bracket symbols in right-hand sides 
have the obvious interpretation of generating a tree structure. 
Obviously, when checking equivalence of two forest transducers, we may
consider their output as \emph{strings}.
Thus, the equivalence problem for deterministic forest transducers is
just a direct instance of the equivalence problem for $y\DT$ transducers.

\begin{corollary}\label{c2}
Equivalence is decidable for deterministic forest transducers.
\end{corollary}

Another, much earlier model of unranked top-down tree transducers
is the \emph{unranked top-down tree transducer}~\cite{Maneth1999}.
Instead of state calls $q(x_i)$ as in an ordinary ranked 
top-down tree transducer, they use calls of the form $L$
where $L$ is a regular language over the set of states $Q$ of
the transducers, plus the special symbol $0$. If the current input node
has $k$-many children, then a word of length $k$ from $L$ is chosen
in order to determine which states translate the children nodes
(where $0$ means that no state translates the corresponding node).
Such a transducer is deterministic, if for every $k$ and every $L$ in
the right-hand side of a rule, $L$ contains at most one string of length $k$.
As an example, consider the unranked top-down tree transducer with the 
following rules:
\[
\begin{array}{lcl}
q_0(a(\cdots))&\to& a(L)\\
q(a(\cdots))&\to& a(L)L
\end{array}
\]
where $L$ is the regular language $q^*$ consting of all strings of the form $qq\cdots q$.
For the input tree $s=a(a(a(a)))$, this transducer first applies the first
rule to obtain $a(q(s_1))$ where $s_1=a(a(a))$. 
We now apply the second rule to obtain $a(a(q(s_2))q(s_2))$ where $s_2=a(a)$. 
Two more applications of the second rule give
$a(a(a(q(a))q(a))a(q(a))q(a))$, and finally we obtain the output tree
$a(a(a(a)a)a(a)a)$.

As mentioned by Perst and Seidl, any unranked top-down tree transducer can
be realized by a forest transducer.
However, they only mention this for nondeterministic transducers.
Thus, given a deterministic unranked top-down tree transducer $M$, 
we can construct an equivalent nondeterministic forest transducer $N$.
It follows from the explanation above that we may consider $N$ as a 
nondeterministic top-down tree-to-string transducer.

By an old result~\cite{DBLP:journals/ipl/Engelfriet78},
for any functional $\DT$ transducer, an equivalent $\DTR$ transducer
can be constructed. Since ``yield'' which turns a tree into its string
of leaf symbols is a function, it directly follows that also for   
any functional $y\DT$ transducer, one can construct an equivalent $y\DTR$ transducer.
Thus, we can construct for $N$ an equivalent $y\DTR$ transducer, for
which equivalence is decidable by Corollary~\ref{cor1}.

\begin{corollary}\label{c3}
Equivalence is decidable for deterministic unranked top-down tree transducers.
\end{corollary}
 
\noindent
Moreover by the constructions from section \ref{s:free}, 
the results stated in in corollaries \ref{c1},\ref{c2} and \ref{c3} also hold when output is not considered in the free monoid, but in the free group.

\section{Conclusion}\label{s:conclusion}

\noindent
We present algorithms for deciding equivalence of deterministic top-down tree-to-string transducers ($y\DT$ transducers).
For $y\DT$ transducers with general output alphabets, 
we provide a construction which encodes outputs over arbitrary 
output alphabets into outputs over a unary alphabet. 
This construction requires to introduce an extra parameter.
For arbitrary $y\DT$ transducers, it results in $y\DMTT$ transducers with unary output alphabet, 
which are non-self-nested whenever the original $y\DT$ transducer is linear.
%
%
For the case of \emph{non-self-nested} unary $y\DMTT$ transducers, 
we show that multi-affine mappings and affine
spaces are sufficient to decide equivalence, whereas in the general case, we had to resort to polynomial ideals.

The key concept which helped us to arrive at a decision procedure in the general case,
are inductive invariants certifying assertions. 
While such invariants can be automatically inferred for monadic input alphabets, 
we were less explicit for non-monadic input alphabets. Here, we only prove 
that an inductive invariant
certifying a polynomial equality \emph{exists}, whenever the equality holds. 
Since enumerating all inductive invariants is rather impractical, 
we presented a more explicit method which allows to systematically construct 
the best inductive invariant up to a given maximal degree.
Together with an explicit enumeration of potential counter-examples,
decidability of equivalence of arbitrary $y\DMTT$ transducers with unary output alphabet follows.
%
%
This result means that \emph{Abelian} equivalence, i.e., equivalence up to the ordering of symbols in the output,
is decidable for general $y\DMTT$ transducers. The same holds true for \emph{growth} equivalence 
where only the lengths of output strings matter.

By means of our simulation of arbitrary output alphabets with unary ones, 
we obtained a randomized polynomial algorithm for deciding in-equivalence of linear $y\DT$ transducers. 
The strongest result, however, is decidability of equivalence of general $y\DT$ transducers 
with arbitrary output alphabets.
Both algorithms are then extended to the case when output is no longer
considered to be in a free monoid, but in a free group.

Still, our decision procedures leave room for generalizations.
%
The equivalence problem for $y\DMTT$ transducers with unary output alphabet being solved,
the equivalence problems (as stated in~\cite{Eng80}) for $y\DMTT$ transducers 
with \emph{arbitrary} output alphabets, and even for $\DMTT$ transducers with tree output,
 remain open.









\bibliographystyle{elsarticle-num} 
\bibliography{literatur}





\end{document}